\documentclass[twocolumn]{aastex62}

\usepackage{amsmath}
\usepackage{longtable}
\usepackage{natbib}

\submitjournal{AAS Journals}
\shorttitle{FeH in Exoplanetary Atmospheres}
\shortauthors{Kesseli et al.}
\pdfoutput=1

\begin{document}

\title{A Search for FeH in Hot-Jupiter Atmospheres with High-Dispersion Spectroscopy}

\correspondingauthor{Aurora Y. Kesseli}
\email{kesseli@strw.leidenuniv.nl}
\author[0000-0002-3239-5989]{Aurora Y. Kesseli}
\affiliation{Leiden Observatory, Leiden University, Postbus 9513, 2300 RA, Leiden, The Netherlands}

\author{I.A.G. Snellen}
\affiliation{Leiden Observatory, Leiden University, Postbus 9513, 2300 RA, Leiden, The Netherlands}

\author{F.J. Alonso-Floriano}
\affiliation{Leiden Observatory, Leiden University, Postbus 9513, 2300 RA, Leiden, The Netherlands}

\author{P. Molli\`ere}
\affiliation{Max-Planck-Institut f\"ur Astronomie, K\"onigstuhl 17, 69117 Heidelberg, Germany}

\author{D.B. Serindag}
\affiliation{Leiden Observatory, Leiden University, Postbus 9513, 2300 RA, Leiden, The Netherlands}

\begin{abstract}

Most of the molecules detected thus far in exoplanet atmospheres, such as water and CO, are present for a large range of pressures and temperatures. In contrast, metal hydrides exist in much more specific regimes of parameter space, and so can be used as probes of atmospheric conditions. Iron hydride (FeH) is a dominant source of opacity in low-mass stars and brown dwarfs, and evidence for its existence in exoplanets has recently been observed at low resolution. We performed a systematic search of archival CARMENES near-infrared data for signatures of FeH during transits of 12 exoplanets. These planets span a large range of equilibrium temperatures (600 $\lesssim T_{eq} \lesssim$ 4000K) and surface gravities (2.5 $\lesssim \mathrm{log} g \lesssim$ 3.5).  We did not find a statistically significant FeH signal in any of the atmospheres, but obtained potential low-confidence signals (SNR$\sim$3) in two planets, WASP-33b and MASCARA-2b. Previous modeling of exoplanet atmospheres indicate that the highest volume mixing ratios (VMRs) of 10$^{-7}$ to 10$^{-9}$ are expected for temperatures between 1800 and 3000K and log $g \gtrsim3$. The two planets for which we find low-confidence signals are in the regime where strong FeH absorption is expected. We performed injection and recovery tests for each planet and determined that FeH would be detected in every planet for VMRs $\geq 10^{-6}$, and could be detected in some planets for VMRs as low as 10$^{-9.5}$. Additional observations are necessary to conclusively detect FeH and assess its role in the temperature structures of hot Jupiter atmospheres.

\vspace{25pt}

\end{abstract}

\section{Introduction}

Hot Jupiters and ultra-hot Jupiters are gas-giant exoplanets that orbit extremely close to their host stars ($\lesssim0.05$AU), and therefore have equilibrium temperatures that are similar to low-mass stars and brown dwarfs (hot Jupiters have 1000 $\lesssim T_{eq} \lesssim$ 2000 K, and ultra-hot Jupiters have 2000 $\lesssim T_{eq} \lesssim$ 4000 K; \citealt{Lothringer2019}). With no similar planets in our own solar system, the atmospheric transmission features in the spectra of transiting hot Jupiters offer the most accessible conditions to learn how planetary atmospheres behave under intense insolation. Studying hot-Jupiter atmospheres has lead to insights on the formation and evolution of these systems \citep{Oberg2011, Mordasini2016, Cridland2019}, atmospheric escape and long-term atmospheric erosion \citep{Vidal2003, Spake2018, Owen2019}, the process of cloud formation \citep{Lecavelier2008, Pont2013, Sing2015, Stevenson2016, Helling2019}, and global atmospheric dynamics \citep{Snellen2010, Kataria2016}. 

Transit transmission spectroscopy of hot and ultra-hot Jupiters has allowed for the detection of many elements and molecules in their atmospheres. The majority of these elements and molecules are present for a wide range of pressures and temperatures, and have thus been detected in both hot and ultra-hot Jupiters, including CO \citep{Snellen2010, Brogi2012, Sheppard2017}, Na \citep{Charbonneau2002, Snellen2008, Wyttenbach2015}, H \citep{Vidal2003, Yan2018, Casasayas2019}, and He \citep{Spake2018, Nortmann2018, Allart2019, Alonso2019b}. While H$_2$O begins to dissociate in ultra-hot Jupiters, it is ubiquitous in exoplanets with temperatures cooler than 2000K \citep{Deming2013, Birkby2013, Sing2016, Sheppard2017}.

Metal oxides (TiO, VO, etc.) and metal hydrides (FeH, MgH, TiH, CaH, etc.) exist within more specific temperature and  pressure ranges and therefore could prove useful as probes of atmospheric conditions \citep[e.g.,][]{Lodders1999}. However, they have proven difficult to conclusively detect even though they are the defining and dominant opacity features in M and L dwarf spectra \citep{Kirkpatrick1999}. Since hot Jupiters have similar equilibrium temperatures as M and L dwarfs, these molecules might be expected in their atmospheres. \citet{Nugroho2017} measured TiO in the atmosphere of WASP-33b at high resolution, but \citet{Herman2020} was not able to reproduce these results. VO has been measured at low-resolution in the atmosphere of WASP-121b \citep{Evans2018}, but has been difficult to confirm at high-resolution due to inaccurate line lists \citep{Merritt2020}. There have also been non-detections of TiO and VO absorption in hot Jupiters with equilibrium temperatures where TiO would be expected, leading to suggestions that TiO and VO are trapped in solids on the much cooler night sides of these exoplanets \citep{Spiegel2009, Sheppard2017}. 

Three studies have reported tentative detections of FeH in four different transiting exoplanets, WASP-62b \citep{Skaf2020}, WASP-79b \citep{Sotzen2020, Skaf2020}, WASP-121b  \citep{Evans2016}, and WASP-127b \citep{Skaf2020}. FeH has also been observed in young directly imaged exoplanets, such as Delorme 1 (AB)b \citep{Eriksson2020}. 
Furthermore, \citet{MacDonald2019} published potential evidence of other metal hydrides (TiH, CrH, and ScH) in HAT-P-26b. However, all of these studies relied on low-resolution spectra, where distinguishing species with overlapping opacities and differentiating them from continuum opacity can be challenging. In addition, the reported signal to noise ratios (SNRs) of the potential signals were all less than five. 

Even though metal oxides and metal hydrides have significantly lower volume mixing ratios (VMRs) than more common molecules and elements, such as CO and H$_2$O, these exotic species can have large effects on exoplanet atmospheres and detecting them can provide important constraints to atmospheric models.  Opacity from TiO and VO are often suggested to be the primary cause of temperature inversions in hot Jupiters \citep[e.g.,][]{Hubeny2003, Fortney2008}. With the debate over the prevalence of TiO and VO, \citet{Lothringer2018} found that opacity from a combination of H$^{-}$, metals, and metal hydrides could produce the required opacity to cause temperature inversions in hot Jupiters without the need for TiO and VO. In addition to being a potential contributor to temperature inversion, FeH also traces weather, cloud formation, and cloud dispersal in L and T dwarfs \citep{Burgasser2008}. 

In this paper, we present results of a systematic search for FeH in the atmospheres of 12 hot gas-giant planets using high dispersion transmission spectroscopy of archival CARMENES data.  The targets cover a range of surface gravities, equilibrium temperatures, and masses, to explore where in parameter space opacity from FeH is important. In Section \ref{s:data} we present the data and discuss the reduction process. In Section \ref{s:model} we introduce the atmospheric models that we used, and in Section \ref{s:signal} we explain how we used these models to retrieve the potential exoplanetary signals. Next, we present the results in Section \ref{s:results} and discuss the expected VMRs of the exoplanets in our study in Section \ref{s:discussion}. Finally, in Section \ref{s:conclusions} we summarize our findings. 

\section{Data and Reduction}
\label{s:data}

\begin{center}
\begin{deluxetable*}{cccccc}[!]
\tabletypesize{\small}
\tablecaption{Transit observations of hot gas giants with CARMENES\label{t:obs}}
\tablehead{
\colhead{Exoplanet} &
\colhead{Observation} & 
\colhead{Number of} & 
\colhead{Phase} & 
\colhead{Exposure Time\tablenotemark{*}} & \colhead{Avg. SNR}
\vspace{-6pt}
\\
\colhead{} &
\colhead{Date} & 
\colhead{Spectra} & 
\colhead{Coverage} & 
\colhead{per Spec. (s)} & 
\colhead{per Spec.}
}
\startdata
KELT-9b & 2017/08/07 & 43 & -$0.078 - 0.100$ & 306 & 92 \\
WASP-33b & 2017/01/05 & 94 & -$0.070 - 0.077$ & 118 & 56 \\
MASCARA-2b & 2017/08/23 & 70 & -$0.031 - 0.034$ & 198 & 91\\
HAT-P-57b & 2018/07/08 & 30 & -$0.045 - 0.049$ & 606 & 51 \\
WASP-76b & 2018/10/03 & 44 & -$0.074 - 0.079$ & 498 & 86 \\
HAT-P-32Ab & 2018/09/01 & 23 & -$0.052 - 0.066$ & 898 & 44 \\
HD 209458b & 2018/09/06 & 91 & -$0.036 - 0.037$ & 198 & 100\\
HD 189733b & 2017/09/07 & 46 & -$0.035 - 0.036$ & 198 & 174 \\
WASP-69b & 2017/08/22 & 35 & -$0.020 - 0.028$ & 398 & 86 \\
WASP-69b & 2017/09/22 & 31 & -$0.022 - 0.020$ & 398 & 75 \\
WASP-107b & 2018/02/24 & 22 & -$0.023 - 0.018$ & 956 & 44 \\
HAT-P-11b & 2017/08/12 & 63 & -$0.024 - 0.044$ & 406 & 107\\
HAT-P-11b & 2017/09/25 & 32 & -$0.019 - 0.018$ & 456 & 97 \\
HAT-P-11b & 2018/07/25 & 28 & -$0.016 - 0.019$ & 498 & 127 \\
GJ 436b & 2017/02/02 & 38 & -$0.270 - 0.025$ & 278 & 85\\
GJ 436b & 2017/02/17 & 36 & -$0.029 - 0.021$ & 278 & 108 \\
GJ 436b & 2018/04/09 & 25 & -$0.012 - 0.023$ & 278 & 50 \\
GJ 436b & 2018/04/16 & 31 & -$0.015 - 0.024$ & 278 & 115 \\
\enddata
\tablenotemark{*}{All exposure times were chosen so that the change in radial velocity of the planet in a single exposure was smaller than the CARMENES NIR pixel size}
\end{deluxetable*}
\end{center}

\begin{splitdeluxetable*}{ccccBcccc}
\tabletypesize{\footnotesize}
\tablecaption{Hot exoplanet system parameters: stellar spectral type, stellar effective temp., stellar radius, system velocity, planet mass, planet radius, planet equilibrium temp., planet log $g$, orbital period, reference mid-transit time, semi-major axis, semi-amplitude of the planet's radial velocity, orbital inclination, planet rotation velocity assuming tidal locking \label{t:planet1}}
\tablehead{
\colhead{} &
\colhead{KELT-9} & 
\colhead{WASP-33} &
\colhead{MASCARA-2} &
\colhead{} & 
\colhead{HAT-P-57} & 
\colhead{WASP-76} &
\colhead{HAT-P-32A} 
}
\startdata
Stellar SpT & B9.5-A0 & A5 & A2 & Stellar SpT & A8 & F7 & late F/early G\\
SpT Ref. & \text{\citealt{Gaudi2017}} & \text{\citealt{Lehman2015}} & \text{\citealt{Talens2018}} & SpT Ref. & \text{\citealt{Hartman2015}} & \text{\citealt{West2016}} & \text{\citealt{Hartman2011}} \\
$T_\mathrm{eff}$ (star; K) & 9600$\pm$400 & 7308$\pm$71 & 8980$\substack{+90\\-130}$ &  $T_\mathrm{eff}$ (star; K) & 6330$\pm$124 & 6250$\pm$100 & 6001$\pm$88 \\
$T_\mathrm{eff}$ Ref. & \text{\citealt{Borsa2019}} & \text{\citealt{Stassun2017}} & \text{\citealt{Talens2018}} & $T_\mathrm{eff}$ Ref. &  \text{\citealt{Stassun2017}} & \text{\citealt{West2016}} & \text{\citealt{Wang2019}} \\
$R_*$ ($R_{\mathrm{Sun}}$) &  $2.418\pm0.058$ & 1.55$\pm$0.05 & 1.60$\pm$0.06 & $R_*$ ($R_{\mathrm{Sun}}$) & 1.538$\substack{+0.0920717\\-0.1053630}$ & 1.73$\pm$0.04 & 1.237$\substack{+0.202\\-0.1}$ \\
$R_*$ Ref. & \text{\citealt{Borsa2019}} & \text{\citealt{Stassun2017}} & \text{\citealt{Talens2018}} & $R_*$ Ref. & \text{\citealt{gaia2018}} & \text{\citealt{West2016}} & \text{\citealt{gaia2018}} \\
$v_{sys}$ (km s$^{-1}$) & -19.819$\pm$0.024 & -3.0$\pm$0.4 & -21.3 $\pm$ 0.4 & $v_{sys}$ (km s$^{-1}$) &  -9.62$\pm$6.87 & -1.0733$\pm$0.0002 & -23.299$\pm$0.0042 \\
$v_{sys}$ Ref. & \text{\citealt{Borsa2019}} & \text{\citealt{Johnson2015}} & \text{\citealt{Talens2018}} & $v_{sys}$ Ref. & \text{\citealt{gaia2018}}& \text{\citealt{Soubiran2018}} & \text{\citealt{Soubiran2018}} \\
$M_{p}$ ($M_{J}$) &  2.88$\pm$0.35 & 2.093$\pm$0.139 & $<$3.382 & $M_{p}$ ($M_{J}$) & 1.41$\pm$1.52	& 0.92$\pm$0.03 & 0.68$\substack{+0.11\\-0.10}$ \\
$M_{p}$ Ref. & \text{\citealt{Borsa2019}} & \text{\citealt{Chakrabarty2019}} & \text{\citealt{Lund2017}} & $M_{p}$ Ref. & \text{\citealt{Stassun2017}} & \text{\citealt{West2016}} & \text{\citealt{Wang2019}} \\
$R_{p}$ ($R_{J}$) &  1.936$\pm$0.047 & 1.60$\pm$0.06 & 1.741$\substack{+0.069\\-0.074}$ & R$_{p}$ ($R_{J}$) & 1.74$\pm$0.36 & 1.83$\substack{+0.06\\-0.04}$ & 1.789$\pm$0.025 \\
$R_{p}$ Ref. & \text{\citealt{Borsa2019}} & \text{\citealt{Stassun2017}} & \text{\citealt{Lund2017}} & $R_{p}$ Ref. & \text{\citealt{Stassun2017}} & \text{\citealt{West2016}} & \text{\citealt{Bonomo2017}} \\
$T_{eq}$ (K) & 4050$\pm$180 & 2781.70$\pm$41.10 & 2260$\pm$50 & $T_{eq}$ (K) & 2200$\pm$76 & 2160$\pm$40 & 1835.7$\substack{+6.8\\-6.9}$ \\
$T_{eq}$ Ref. & \text{\citealt{Gaudi2017}} & \text{\citealt{Chakrabarty2019}} & \text{\citealt{Talens2018}} & $T_{eq}$ Ref. & \text{\citealt{Hartman2015}} & \text{\citealt{West2016}} & \text{\citealt{Wang2019}} \\
log $g$ $^\dagger$ & 3.28 & 3.4 & $<$3.46 & log $g$ $^\dagger$ & 3.08 & 2.85 & 2.76 \\
$P_{orb}$ (d) & 1.4811235 & 1.21986983 & 3.4741070 & $P_{orb}$ (d) & 2.4653 & 1.809886 & 2.1500082 \\
$P$ Ref. & \text{\citealt{Gaudi2017}} & \text{\citet{Stassun2017}} & \text{\citealt{Lund2017}} & $P$ Ref. & \text{\citealt{Stassun2017}} & \text{\citealt{West2016}} & \text{\citealt{Wang2019}} \\
$T_0$ (d) & 2457095.68572 & 2452984.82964 & 2457909.5906 & $T_0$ (d) & 2455113.48127 & 2456107.85507 & 2455867.402743 \\
$T_0$ Ref. & \text{\citealt{Gaudi2017}} & \text{\citealt{Turner2016}} & \text{\citealt{Talens2018}} & $T_0$ Ref. & \text{\citealt{Hartman2015}} & \text{\citealt{West2016}} & \text{\citealt{Wang2019}} \\
$a$ (au) & 0.03462$\substack{+0.00110\\-0.00093}$ & 0.0259$\pm$0.0005 & 0.0542$\substack{+0.0014\\-0.0021}$ & $a$ (au) & 0.0406$\pm$0.0011 & 0.0330$\pm$0.0005& 0.03427$\substack{+0.00040\\-0.00042}$ \\
$a$ Ref. & \text{\citealt{Gaudi2017}} & \text{\citealt{Turner2016}} & \text{\citealt{Lund2017}}& $a$ Ref. & \text{\citealt{Hartman2015}} & \text{\citealt{West2016}} & \text{\citealt{Bonomo2017}} \\
$K_p$ (km s$^{-1}$)$^\dagger$ & 269 & 231 & 170 & $K_p$ (km s$^{-1}$)$^\dagger$ & 180 & 198 & 172 \\
$i$ (deg) & 86.79$\pm$0.25 & 86.63 & 86.12 & $i$ (deg) & 88.26$\pm$0.85 & 88.0$\substack{+1.3\\-1.6}$ & 88.90$\pm$0.40\\
$i$ Ref. & \text{\citealt{Gaudi2017}} & \text{\citealt{Chakrabarty2019}} & \text{\citealt{Lund2017}} & $i$ Ref. & \text{\citealt{Hartman2015}} & \text{\citealt{West2016}} & \text{\citealt{Stassun2017}} \\
$v_{rot}$ (km s$^{-1}$)$^\dagger$ & 6.6 & 6.7 & 2.5 & $v_{rot}$ ((km s$^{-1}$)$^\dagger$ & 3.6 & 5.1 & 4.2  \\
\enddata
\end{splitdeluxetable*}

\setcounter{table}{1}
\begin{splitdeluxetable*}{ccccBcccc}
\tabletypesize{\footnotesize}
\tablecaption{\textbf{cont.} \label{t:planet2}}
\tablehead{
\colhead{} &
\colhead{HD 209458} &
\colhead{HD 189733} &
\colhead{WASP-69} &
\colhead{} &
\colhead{WASP-107} & 
\colhead{HAT-P-11} & 
\colhead{GJ 436} 
}
\startdata
Stellar SpT & G0 & K0-2 & K5 & Stellar SpT & K6 & K4 & M2.5 \\
SpT Ref. & \text{\citealt{delBurgo2016}} & \text{\citealt{Salz2015}} & \text{\citealt{Anderson2014}} & SpT Ref. & \text{\citealt{Anderson2017}} & \text{\citealt{Bakos2010}} & \text{\citealt{Butler2004}}\\
$T_\mathrm{eff}$ (star; K) & 6091$\pm$10 & 5052$\pm$16 & 4700$\pm$50 & $T_\mathrm{eff}$ (star; K) & 4430$\pm$120 & 4708$\pm$84 & 3479$\pm$60 \\
$T_\mathrm{eff}$ Ref. & \text{\citealt{Stassun2017}} & \text{\citealt{Stassun2017}} & \text{\citealt{Stassun2017}} & $T_\mathrm{eff}$ Ref.& \text{\citealt{Anderson2017}} & \text{\citealt{Stassun2017}} & \text{\citealt{Bourrier2018}} \\
$R_*$ ($R_{\mathrm{Sun}}$) & 1.19$\pm$0.02 & 0.75$\pm$0.01 & 0.818$\substack{+0.02\\-0.027}$ & $R_*$ ($R_{\mathrm{Sun}}$) & 0.66$\pm$0.02 & 0.683$\pm$0.009 & 0.449$\pm$0.019 \\
$R_*$ Ref. & \text{\citealt{Stassun2017}} & \text{\citealt{Stassun2017}} & \text{\citealt{gaia2018}} & $R_*$ Ref. & \text{\citealt{Anderson2017}} & \text{\citealt{Yee2018}} & \text{\citealt{Bourrier2018}} \\
$v_{sys}$ (km s$^{-1}$) & -14.743$\pm$0.0012 & -2.317$\pm$0.0009 & -9.37$\pm$0.21 &  $v_{sys}$ (km s$^{-1}$) & 13.74$\pm$0.31 & -63.452$\pm$0.0011 & 9.609$\pm$0.0010\\
$v_{sys}$ Ref. & \text{\citealt{Soubiran2018}} & \text{\citealt{Soubiran2018}} & \text{\citealt{gaia2018}} & $v_{sys}$ Ref. & \text{\citealt{gaia2018}} & \text{\citealt{Soubiran2018}} & \text{\citealt{Soubiran2018}} \\
$M_{p}$ ($M_{J}$) & 0.73$\pm$0.04 & 1.123$\pm$0.045& 0.2600$\pm$0.0185 &  $M_{p}$ ($M_{J}$) & 0.12$\pm$0.01& 0.09$\pm$0.01 & 0.0728$\pm$0.0024\\
$M_{p}$ Ref. & \text{\citealt{Stassun2017}} & \text{\citealt{Bonomo2017}} & \text{\citealt{Casasayas2017}} &  $M_{p}$ Ref. & \text{\citealt{Anderson2017}} & \text{\citealt{Stassun2017}} & \text{\citealt{Turner2016}} \\
$R_{p}$ ($R_{J}$) & 1.39$\pm$0.02 & 1.13$\pm$0.01 & 1.057$\pm$0.017 & $R_{p}$ ($R_{J}$) & 0.94$\pm$0.02 & 0.389$\pm$0.005 & 0.3739$\pm$0.0097\\
$R_{p}$ Ref. & \text{\citealt{Stassun2017}} & \text{\citealt{Stassun2017}} & \text{\citealt{Casasayas2017}} & $R_{p}$ Ref. & \text{\citealt{Anderson2017}} & \text{\citealt{Yee2018}} & \text{\citealt{Turner2016}} \\
$T_{eq}$ (K) & 1450 & 1200 & 963$\pm$18 	
& $T_{eq}$ (K) & 770$\pm$60 & 878$\pm$15 & 686$\pm$10\\
$T_{eq}$ Ref. & \text{\citealt{Sing2016}} & \text{\citealt{Sing2016}} & \text{\citealt{Anderson2014}} & $T_{eq}$ Ref. & \text{\citealt{Anderson2017}} & \text{\citealt{Bakos2010}} & \text{\citealt{Turner2016}} \\
log $g$ $^\dagger$ & 2.99 & 3.36 & 2.74 & log $g$ $^\dagger$ & 2.55 & 3.16 & 3.12 \\
$P_{orb}$ (d) & 3.52474859 & 2.21857567 & 3.868140$\pm$0.000002 & $P_{orb}$ (d) &  5.721490$\pm$0.000002 & 4.887820$\pm$0.000007& 2.64389803$\pm$0.00000026\\
$P$ Ref. & \text{\citealt{Stassun2017}} & \text{\citealt{Stassun2017}} & \text{\citealt{Stassun2017}} & $P$ Ref. & \text{\citealt{Anderson2017}} & \text{\citealt{Stassun2017}} & \text{\citealt{Bourrier2018}} \\
$T_0$ (d) & 2452826.629283 & 2454279.436714 & 2455748.83422 & $T_0$ (d) &  2456514.4106 & 2454957.812464 & 2454865.084034\\
$T_0$ Ref. & \text{\citealt{Bonomo2017}} & \text{\citealt{Agol2010}} & \text{\citealt{Bonomo2017}} & $T_0$ Ref. & \text{\citealt{Anderson2017}} & \text{\citealt{Sanchis2011}} & \text{\citealt{Bourrier2018}}\\
$a$ (au) & 0.04707$\substack{+0.00045\\-0.00047}$ & 0.03100$\substack{+0.00059\\-0.00061}$ & 0.04527$\substack{+0.00053 \\ -0.00054}$& $a$ (au) & 0.055$\pm$0.001& 0.05254$\substack{+0.00064\\-0.00066}$ & 0.03109$\pm$0.00074\\
$a$ Ref. & \text{\citealt{Bonomo2017}}  & \text{\citealt{Bonomo2017}} & \text{\citealt{Bonomo2017}} & $a$ Ref. & \text{\citealt{Anderson2017}} & \text{\citealt{Yee2018}} & \text{\citealt{Turner2016}} \\
$K_p$ (km s$^{-1}$) $^\dagger$ & 145 &  153 & 127 & $K_p$ (km s$^{-1}$) $^\dagger$ & 105 & 117 &
128 \\
$i$ (deg) & 86.71$\pm$0.05 & 85.580$pm$0.060 & 86.71$\pm$0.20 & $i$ (deg) & 89.7$\pm$0.2 & 88.50$\pm$0.60 & 86.774$\pm$0.030\\
$i$ Ref. & \text{\citealt{Stassun2017}} & \text{\citealt{Bonomo2017}} & \text{\citealt{Anderson2014}} & $i$ Ref. & \text{\citealt{Anderson2017}} & \text{\citealt{Stassun2017}} &  \text{\citealt{Turner2016}} \\
$v_{rot}$ ((km s$^{-1}$)$^\dagger$ & 2.0 & 2.6 & 1.4 & $v_{rot}$ ((km s$^{-1}$)$^\dagger$ & 0.8 & 0.4 & 0.7 \\
\enddata
\tablenotemark{$^\dagger$}{calculated using the listed parameters}\\
\end{splitdeluxetable*}

CARMENES is a high resolution echelle spectrograph \citep{Quirrenbach2018}, which is capable of characterizing atmospheres of transiting exoplanets \citep[e.g.,][]{Alonso2019a}. CARMENES is installed on the 3.5 meter telescope at the Calar Alto Observatory and contains a near-infrared channel (NIR; R$\sim$80,000) and an optical channel (VIS; R$\sim$95,000). In this study, we used data from the NIR channel, which covers y, J, and most of H band, spanning wavelengths $0.96 < \lambda < 1.71 \mu m$. 
We downloaded all of the transits of hot gas-giant planets that were publicly available on the Calar Alto Public Archives (CAHA Archive\footnote{\url{http://caha.sdc.cab.inta-csic.es/calto/jsp/searchform.jsp}}) through November of 2019, and exclude those transits that were severely affected by weather conditions like clouds or high humidity.
Table \ref{t:obs} summarizes the observations of each exoplanet transit.

The exoplanetary systems cover a large range of parameter space in terms of planet mass ($0.07 \leq \mathrm{M}_p \leq 3.38\ \mathrm{M}_{J}$), radius ($0.37 \leq \mathrm{R}_p \leq 1.89\ \mathrm{R}_{J}$), surface gravity ($2.74 \leq \mathrm{log}\ g \leq 3.46$), equilibrium temperature ($686 \leq \mathrm{T}_{eq} \leq 4050$ K), and host star spectral type (B9.5 through M2.5). Table \ref{t:planet1} shows a compiled list of all of these parameters, as well as other important properties of the systems that are used in the following analysis. 

\begin{figure*}
\begin{center}
\includegraphics[width=\linewidth]{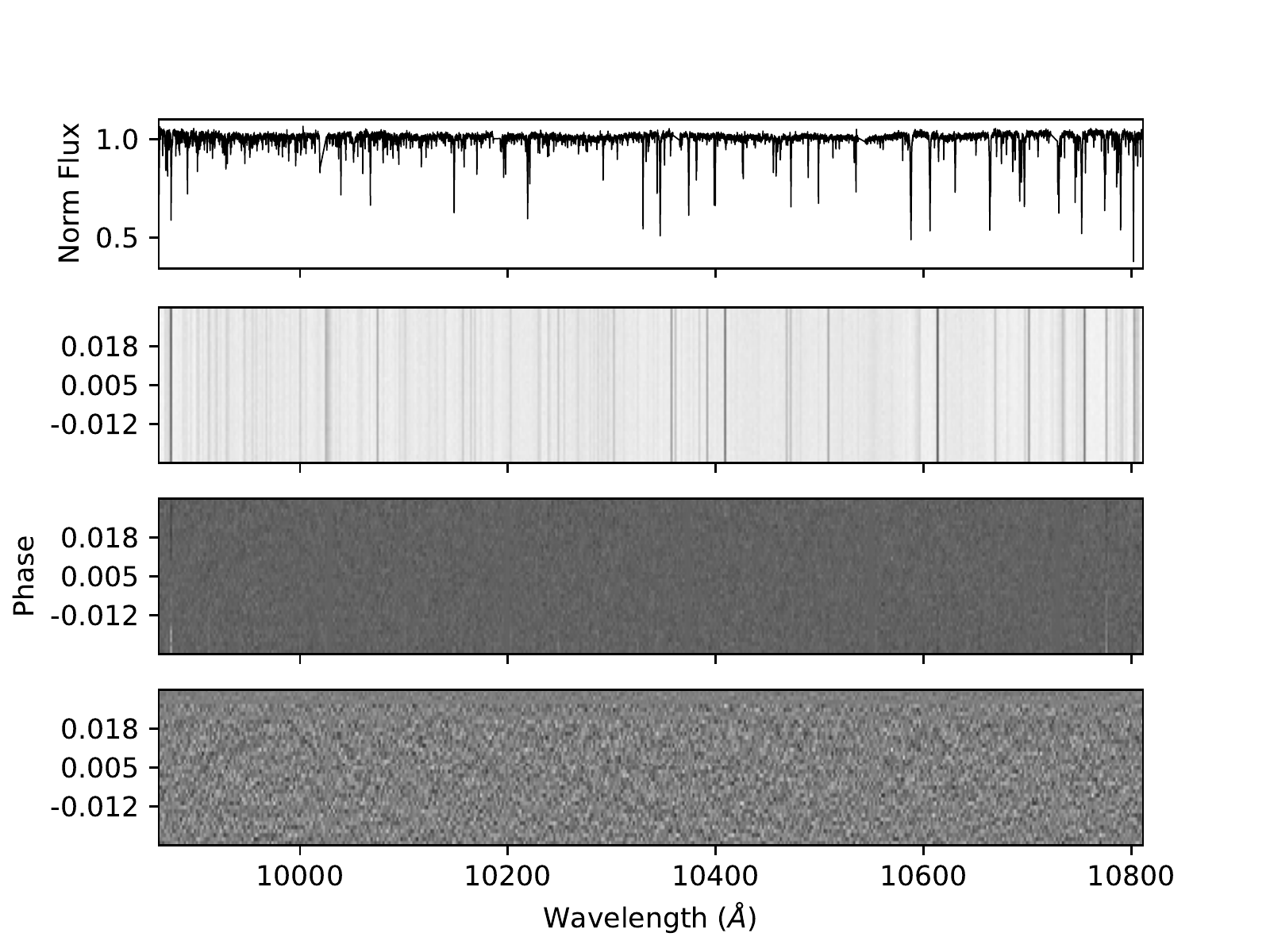}
\caption{\small 
Example of the analysis steps of the HD 189733 system. The top plot shows one spectrum that has been order-combined and flattened and contains the full wavelength range that we used in our analysis. Next, we show all 46 of the sigma-clipped and normalized spectra in grayscale. The dark vertical lines are the stellar absorption features mixed with a few telluric absorption lines. The third panel shows the data after the stellar lines have been removed and another high-pass filter has been applied. The dominant variations left are due to telluric lines mostly at the edges of the spectra. The bottom panel shows the final spectra after 9 \texttt{SYSREM} iterations have been applied. The spectra appear uniform and lack any significant trends in time. }
\label{f:SpecReduction}
\end{center}
\end{figure*}

The downloaded data were already reduced using the CARACAL pipeline \citep{Caballero2016}. The pipeline performs a bias and flat field correction as well as a wavelength calibration. The wavelength solution is given in vacuum wavelengths in Earth's rest frame. The pipeline does not perform any telluric correction, and any telluric absorption or sky emission that is present in the spectra is removed later during our analysis process.  The region where the main FeH bandhead exhibits a peak opacity ($\sim$0.99 $\mu$m) is directly between water bands so there is minimal telluric contamination to begin with. As telluric correction is a major hurtle in analyzing ground-based exoplanet observations in the NIR, the position of the FeH bandhead allows us to efficiently analyze transits from many planets. 

In order to uncover the small planetary transit signal, we had to further analyze the data and remove the stellar absorption features and any remnant noise. We followed a similar set of steps as \citet{Alonso2019a} and \citet{SanchezLopez2019}, except for our treatment of the spectral orders. In previous studies, the orders were handled separately until the very end, when the final 1D cross correlation functions were combined. This can decrease the significance of the signal if orders with less strong absorption are simply added to orders of strong absorption, or can lead to a falsely inflated signal if the orders are weighted before they are combined. Instead, we combined the orders into a single spectrum from the start (see Section \ref{s:model} to see which orders are included). This simplified and reduced the computational time of the process, allowing us to efficiently analyze many transits, and did not significantly change the final cross correlation function. To combine the orders, we first normalized each order by fitting a third degree polynomial to the continuum and then interpolated each order onto a wavelength grid that was sampled uniformly in log wavelength space in 0.2 km s$^{-1}$ increments. This spacing is highly oversampled as the CARMENES resolution is about 3 km s$^{-1}$, but resulted in virtually no information being lost in the interpolation process and does not cause any problems during the later analysis steps. Any regions that overlapped between the orders were combined together using an average that was weighted by the uncertainty in each pixel. An example of the single order-combined spectrum is shown in the top panel of Figure \ref{f:SpecReduction}.

Figure \ref{f:SpecReduction} outlines the analysis steps for HD 189733 as an example of our process. To begin, we removed any 5-sigma outliers from cosmic rays or bad pixels that were missed by the automatic reduction pipeline (second panel of Figure \ref{f:SpecReduction}). Next, we removed the stellar absorption features. Since in every case the star's radial velocity changes by less than the width of a CARMENES pixel over the course of our observations, while that of the planet is rapidly increasing at up to 10 km s$^{-1}$ per hour, we created a time averaged spectrum and then divided each time series spectrum by this average. This process removed any signal that was constant in wavelength over the observation times, but left any signal that was not constant in time (i.e. any planet absorption).
Before this step could be completed, however, we corrected for the change in the stellar line positions over the course of the observations due to the motion of the Earth (barycentric velocity correction). To calculate the barycentric velocity of each exposure, we used the \texttt{barycorr} online application to convert the mid exposure MJD to a barycentric velocity correction with a precision of 3 m s$^{-1}$ \citep{Wright2014}. 

We next applied a high-pass filter with a width of 1000 pixels (200 km s$^{-1}$) and again performed a 5-sigma clipping in case any overall shape differences or outliers remained in the spectra. Finally, we divided by the standard deviation of each pixel in time, which effectively down-weighted the wavelength regions with large standard deviations due to tellurics, bad pixels or cosmic rays.

At this point the dominant remaining features were due to telluric contamination (see third panel in Figure \ref{f:SpecReduction}). To remove these features, the strong lines are often masked and any residuals are removed with the \texttt{SYSREM} algorithm. We experimented with masking and \texttt{SYSREM}, but chose to only use \texttt{SYSREM} in the following results as the telluric features are not strong in our wavelength region and \citet{Cabot2019} suggested that masking regions of the spectrum could contribute to spurious high SNR peaks. \texttt{SYSREM} iteratively performs principle component analysis, allowing for unequal uncertainties at each wavelength point, to remove systematic trends in photometric or spectroscopic data due to trends in temperature, airmass, and more \citep{Tamuz2005, Mazeh2007}.  Numerous studies have thoroughly tested and validated the use of \texttt{SYSREM} for removing telluric signals in high resolution spectroscopy \citep[e.g.,][]{Birkby2017, Nugroho2017, Cabot2019}. Even with the extensive testing, \texttt{SYSREM} cannot be applied blindly to the spectra, as it can remove the planetary signal if too many iterations are applied. To find the ideal number of \texttt{SYSREM} iterations, we tried a range of values and observed how the SNR evolved over each iteration. Section \ref{s:injections} shows the results of these tests on the injected signal. 

\section{Atmospheric Transmission Model} 
\label{s:model}

\begin{center}
\begin{deluxetable*}{cccc}
\tabletypesize{\small}
\tablecaption{Planet atmospheric model parameters\label{t:model}}
\tablehead{
\colhead{Exoplanet} &
\colhead{PT profile} & 
\colhead{H$^-$ VMR (0.01 bars)} & 
\colhead{Cloud Base} \\
}
\startdata
KELT 9b & 4000 K\tablenotemark{1} & $1.0\times 10^{-10}$ & none, \textbf{0.1 bar}\tablenotemark{2}, 0.01\\
WASP 33b & \citealt{Haynes2015} & $1.3\times 10^{-10}$ &  none, \textbf{0.1 bar}, 0.01\\
MASCARA 2b & 2250 K & $3.1 \times10^{-11}$ & none, \textbf{0.1 bar}, 0.01 \\
HAT-P-57b & 2200 K & $2.2 \times10^{-11}$ & none, \textbf{0.1 bar}, 0.01 \\
WASP 76b & 2150 K  & $1.5 \times10^{-11}$  & none, \textbf{0.1 bar}, 0.01 \\
HAT-P-32Ab & \citealt{Nikolov2018} & $1.0 \times10^{-11}$ &  none, \textbf{0.1 bar}, 0.01, 0.001, 0.0001 bar\tablenotemark{3}\\
HD 209458b & \citealt{Brogi2017} & $2.0 \times10^{-16}$ & none, \textbf{0.1 bar}, 0.01 bar\\
HD 189733b & 1200 K & $3.7 \times10^{-18}$ & none, \textbf{0.1 bar}, 0.01 bar\\
WASP 69b & 1000 K & $3.2 \times10^{-21}$ &  none, \textbf{0.1 bar}, 0.01 bar\\
WASP 107b & 1000 K & $3.2 \times10^{-21}$ & none, \textbf{0.1 bar}, 0.01 bar\\
HAT-P-11b & 1000 K & $3.2 \times10^{-21}$ & none, \textbf{0.1 bar}, 0.01 bar\\
GJ 436b & 1000 K & $3.2 \times10^{-21}$ & none, \textbf{0.1 bar}, 0.01 bar\\
\enddata
\tablenotemark{1}{When no published PT profile was available we assumed an isothermal profile at the equilibrium temperature}\\
\tablenotemark{2}{Bolded values are used in the fiducial models and for most of the analysis}\\
\tablenotemark{3}{Based on the literature, it is unclear if HAT-P-32 has a high cloud deck or not \citep{Damiano2017, Nortmann2016}, so we explored a larger area of parameter space }\\
\end{deluxetable*}
\end{center}

\begin{figure}[h!]
\begin{center}
\includegraphics[width=\linewidth]{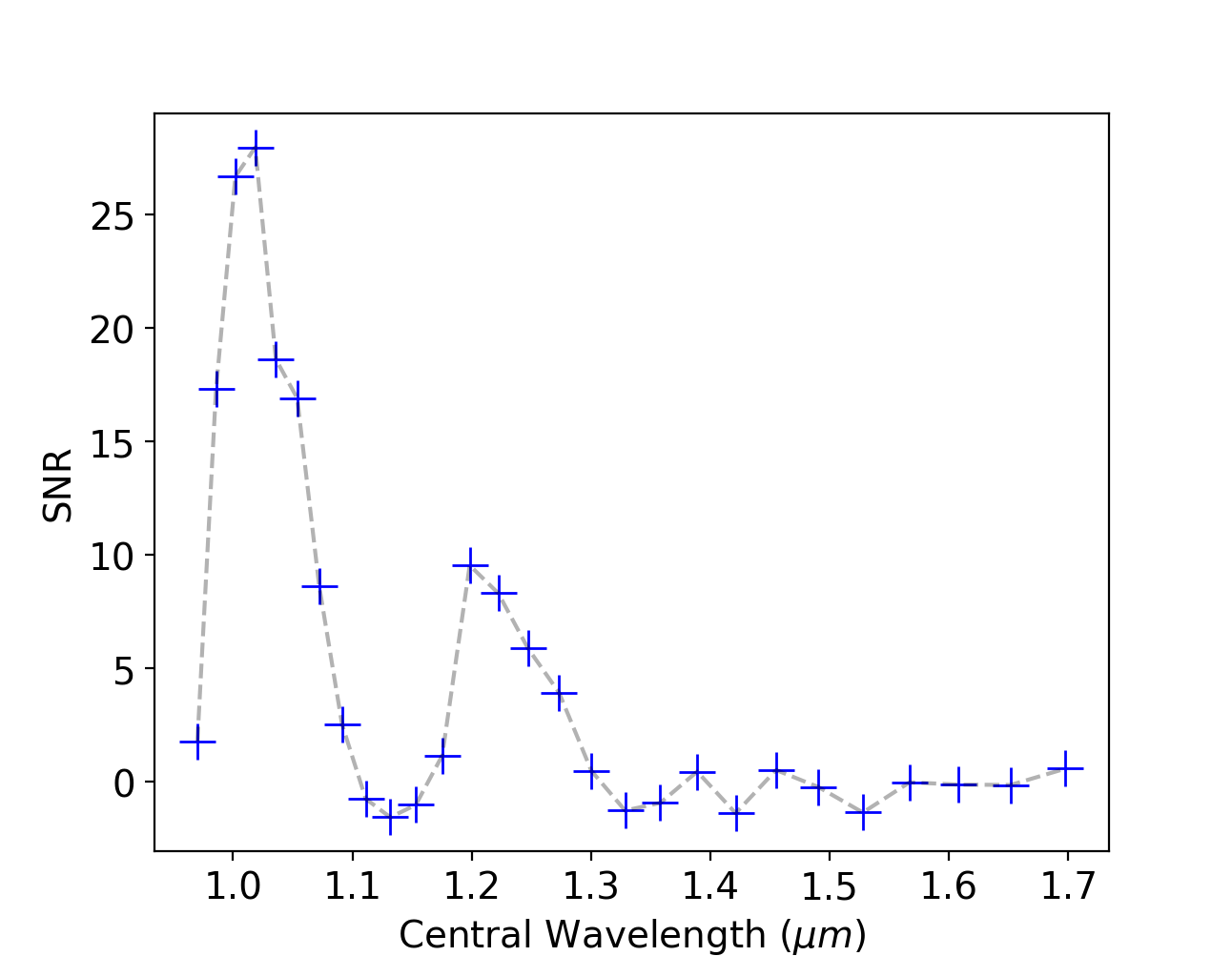}
\caption{\small 
The cross correlation signal between the ExoMol line list and Teegarden's star for each CARMENES NIR spectral order. The main FeH bandhead begins at 0.99 $\mu$m. Water bands surround the FeH bandhead and are the cause of the decrease in SNR around 0.9, 1.15, and 1.3 $\mu$m. A second peak in SNR occurs around 1.25$\mu$m, but it is about a third as strong, and there is significantly more telluric contamination in this wavelength region than around the primary peak. }
\label{f:Carmenes_dM_FeH}
\end{center}
\end{figure}

\begin{figure*}
\begin{center}
\includegraphics[width=0.48\linewidth]{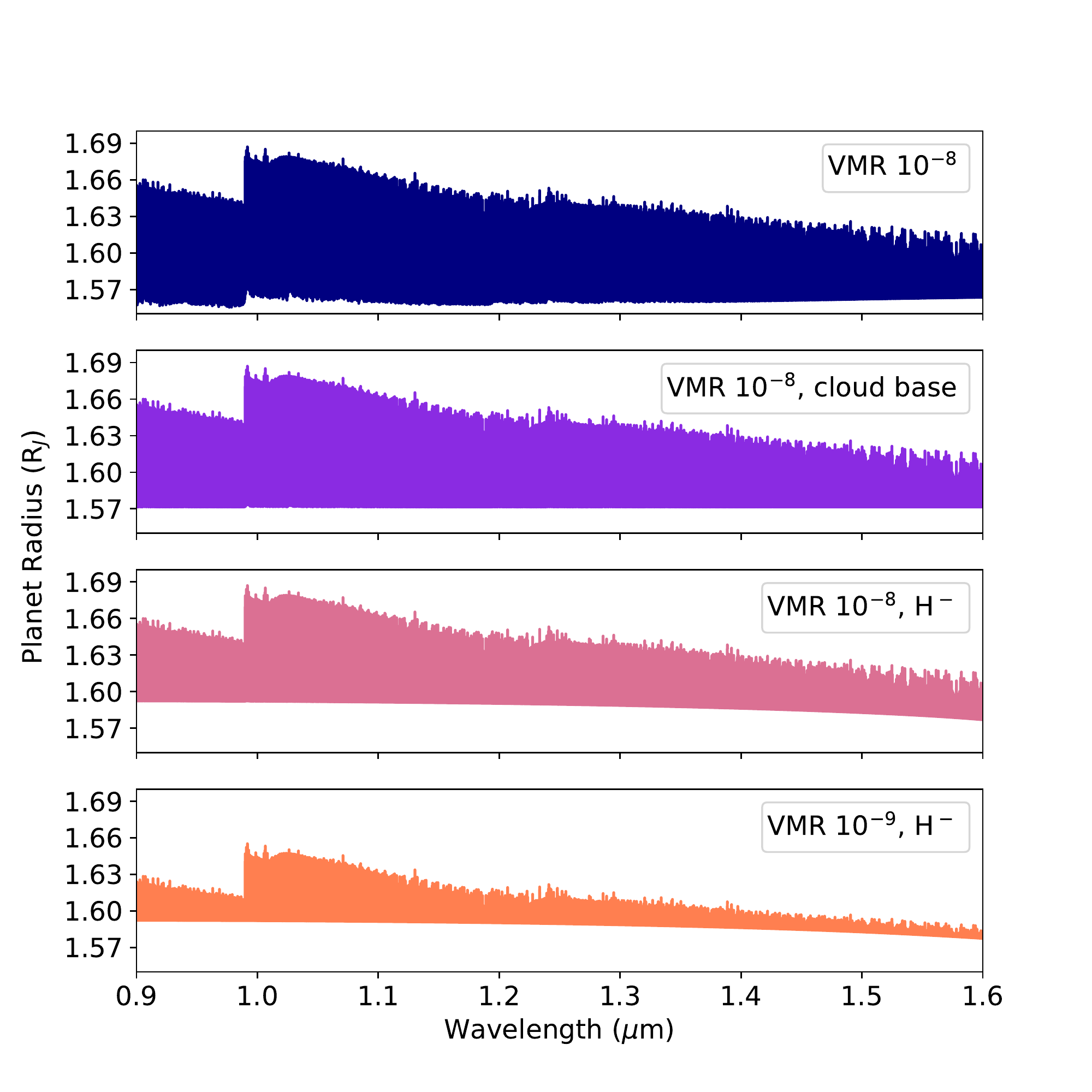}
\includegraphics[width=0.48\linewidth]{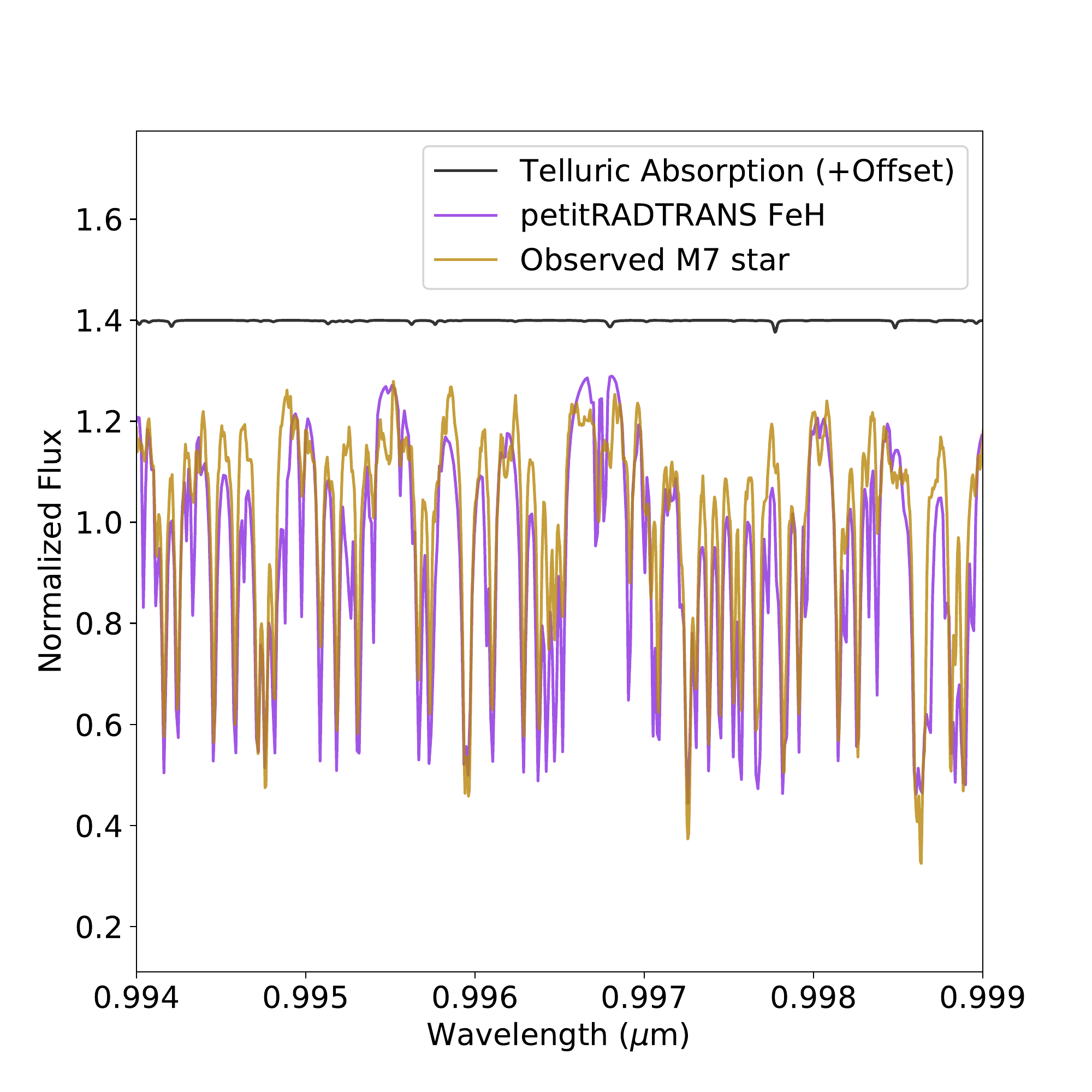}
\caption{\small 
Left: Example of the atmospheric models from petitRADTRANS for WASP-33b for different continuum opacity sources and FeH VMRs. All of the models include opacity from H$_{2}-$H$_{2}$ and H$_{2}-$He collisions, while in the bottom three plots other sources of continuum opacity are also included. In the second plot we have included a cloud base at 0.1 bar, below which the atmosphere cannot be probed. In the bottom two plots, the continuum is set by opacity from H$^-$ (with a VMR of 10$^{-10}$). As expected, by adding more continuum opacity sources, the line strengths are decreased. The bottom plot shows that by decreasing the VMR of FeH, the line strengths also decrease. Right: Enlarged version of the second FeH model showing about 5\% of the wavelength coverage we used, as well as a non-telluric corrected CARMENES spectrum of Teegarden's star and the telluric absorption model for this spectrum from molecfit \citep{Smette2015}. Some water lines are present but they are much weaker and sparser than the FeH lines. The model and observations are in very good agreement and there are thousands of FeH lines densely spaced throughout this wavelength region.  }
\label{f:pRT}
\end{center}
\end{figure*}

We created a model 
transmission spectrum of FeH for each exoplanet using the petitRADTRANS package \citep{Molliere2019} to test whether FeH can be observed in the atmospheres of these planets.
petitRADTRANS is a radiative transfer code, designed specifically for spectral characterization of exoplanetary atmospheres. The code takes as input a temperature-pressure profile, the planetary radius, the surface gravity, the relative abundances of the requested species, and the mean molecular weight of the atmosphere, and produces a transmission or emission spectrum at low or high spectral resolution. 
The relative abundances of the species are required to be in units of mass fractions, and not VMRs, so we multiply by the molecular weight over the mean molecular weight to convert to mass fractions, where applicable.

A pre-computed opacity line list of FeH is available with the code. The opacity line list of FeH is sourced from the ExoMol library and uses the empirically determined FeH lines from \citet{Wende2010}. We tested this line list to ensure its accuracy by comparing the list to a high SNR CARMENES spectrum of Teegarden's star, an M dwarf with a similar effective temperature (2700 K) to many of the exoplanets in our study. To accurately compare the two, we used petitRADTRANS to make a model of FeH using the parameters of Teegarden's star (shown in the right panel of Figure \ref{f:pRT}). We cross correlated each spectral order separately with the ExoMol line list to determine in which orders FeH was detectable and at what level.  Figure \ref{f:Carmenes_dM_FeH} shows that FeH is strongly detected in the six orders around the primary bandhead. There is another peak in the SNR function around 1.25 $\mu$m, but this region is heavily contaminated by telluric water and oxygen lines and we find when these orders are added to our injection and retrieval tests (see Section \ref{s:injections}), the SNR of our retrieved signal is actually decreased. We therefore use only the 6 orders that span wavelengths 0.98 $-$ 1.08 $\mu$m.

With the validated line list, we created transmission spectra of FeH using the high-resolution mode (R$= 10^6$) of petitRADTRANS (see Figure \ref{f:pRT}). The planetary radii and surface gravities that were used as input are all listed in Table \ref{t:planet1}. When available, we used published pressure-temperature (PT) profiles (see Table \ref{t:model}). If no PT profile was available, we assumed an isothermal profile at the planet's equilibrium temperature. Transmission spectra are not very sensitive to small changes in the PT profile and so this should be sufficient for our purpose. To determine the mean molecular weight (MMW) of each planet, we implemented a simple equilibrium chemistry code to minimize Gibbs free energy. The code is described in Appendix A2 of \citet{Molliere2017}, and takes the PT profile, the C/O and Fe/H ratios (both assumed to be solar) as input, and returns the MMW and other elemental or molecular abundances at each pressure in the atmosphere.

We also included continuum opacity from H$_{2}-$H$_{2}$ collisions, H$_{2}-$He collision, and H$^{-}$ opacity in all of the atmospheres. Many recent works have emphasized the importance of including H$^{-}$ opacity in hot exoplanet atmospheres \citep[e.g., ][]{Freedman2014, Lothringer2018, Arcangeli2018}. \citet{Freedman2014} showed that H$^{-}$ was the dominant continuum opacity source around one micron for an atmosphere of 2600 K. 
The abundances of H$^{-}$, free electrons, and H are included in the equilibrium chemical modeling. 
In some models we also experimented with including a cloud base, below which the atmosphere cannot be probed. Table \ref{t:model} show the continuum opacity sources explored in the models, and Figure \ref{f:pRT} shows the effect of these changes to the continuum opacity on the resulting model atmospheres. 

Finally, we tested a range of different VMRs for FeH, from $10^{-4}$ to $10^{-10}$. We did not utilize any chemical modeling to calculate the VMRs of FeH at different altitudes, but instead kept the VMR of FeH constant. This choice was motivated in part to facilitate comparisons with previous studies, which all used constant VMRs of FeH. It was also motivated by discrepancies between observations of brown dwarfs and chemical models; many low-temperature brown dwarfs show evidence for FeH even though the models predict that FeH should have condensed out of the atmosphere (see Section \ref{s:discussion} for more details). Even though a constant VMR is less realistic than a VMR that changes with altitude, the resulting cross correlation functions are not significantly affected because transmission spectra are not very sensitive to these changes. 

To make the models more realistic, we took into consideration the rotation of the planet in a method similar to \citet{Brogi2016}, assuming tidally locked planets. The majority of the planets have rotation velocities that are smaller than the resolution element of CARMENES (see $v_{rot}$ in Table \ref{t:planet1}), but for the fastest rotating planets this effect is significant. We also reduced the resolution of the models to match the CARMENES spectrograph (R$\sim$80,000) by convolving the models with a Gaussian kernel, and then interpolated the models onto the same wavelength grid as the data. Before cross correlation of the models with the data, we applied the same high-pass filter to the models as was used on the data.

\section{Signal Retrieval}
\label{s:signal}

\subsection{Cross Correlation \label{s:cc}}

\begin{figure}
\begin{center}
\includegraphics[width=\linewidth]{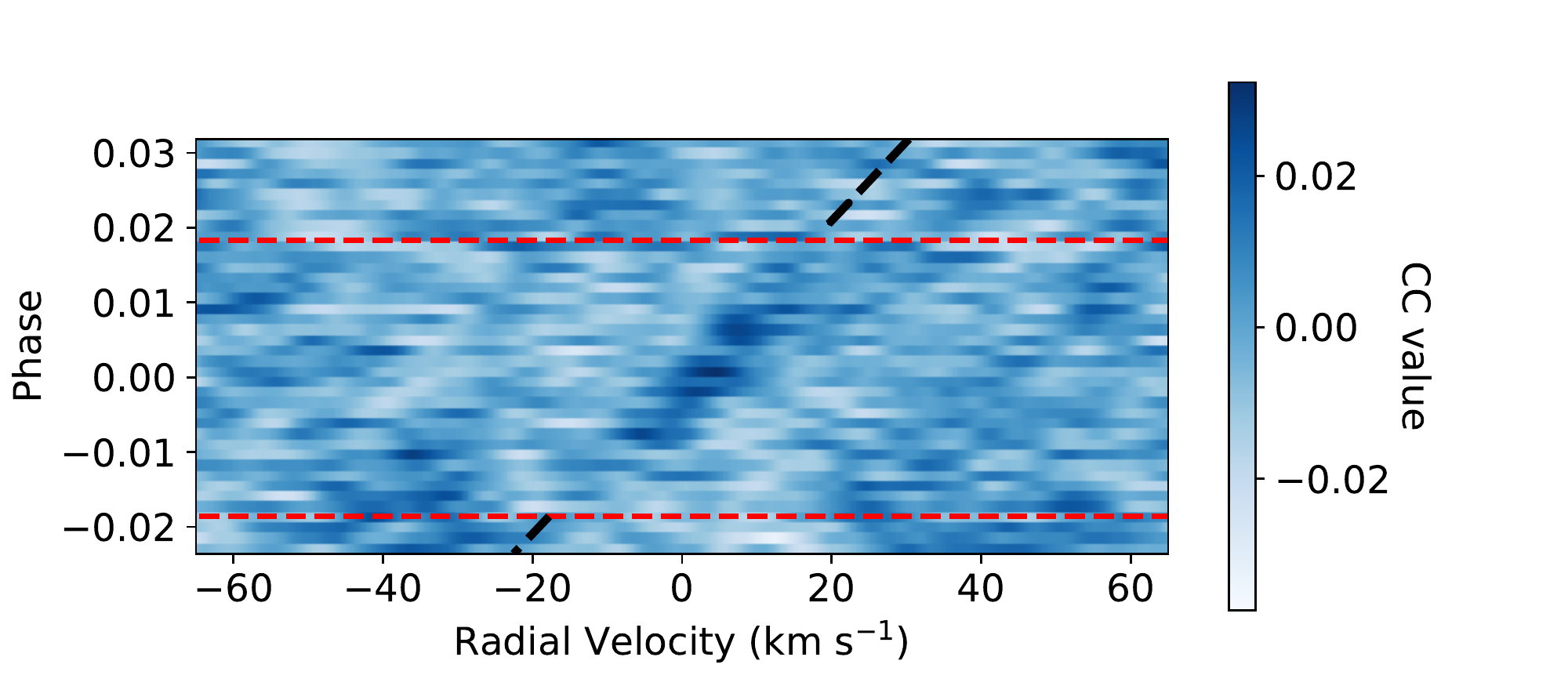}
\includegraphics[width=\linewidth]{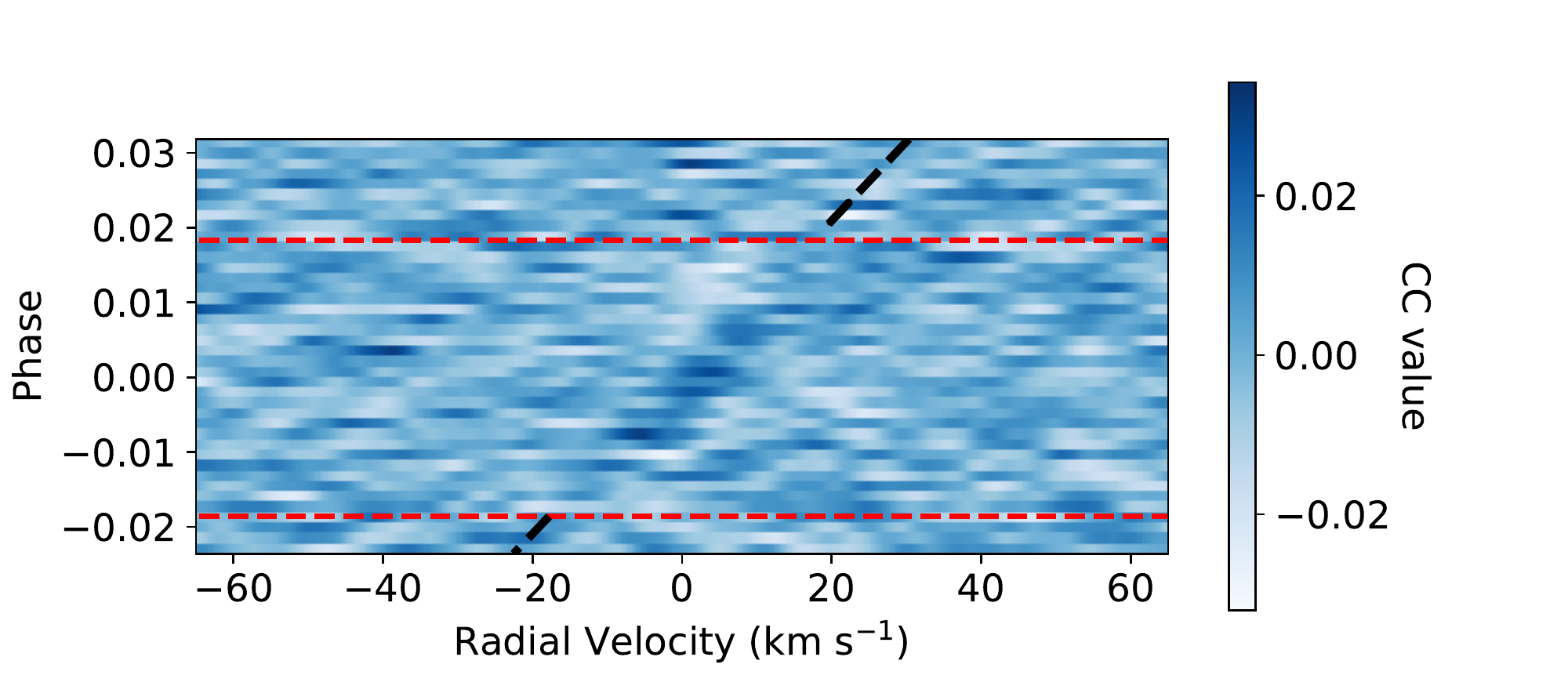}
\includegraphics[width=\linewidth]{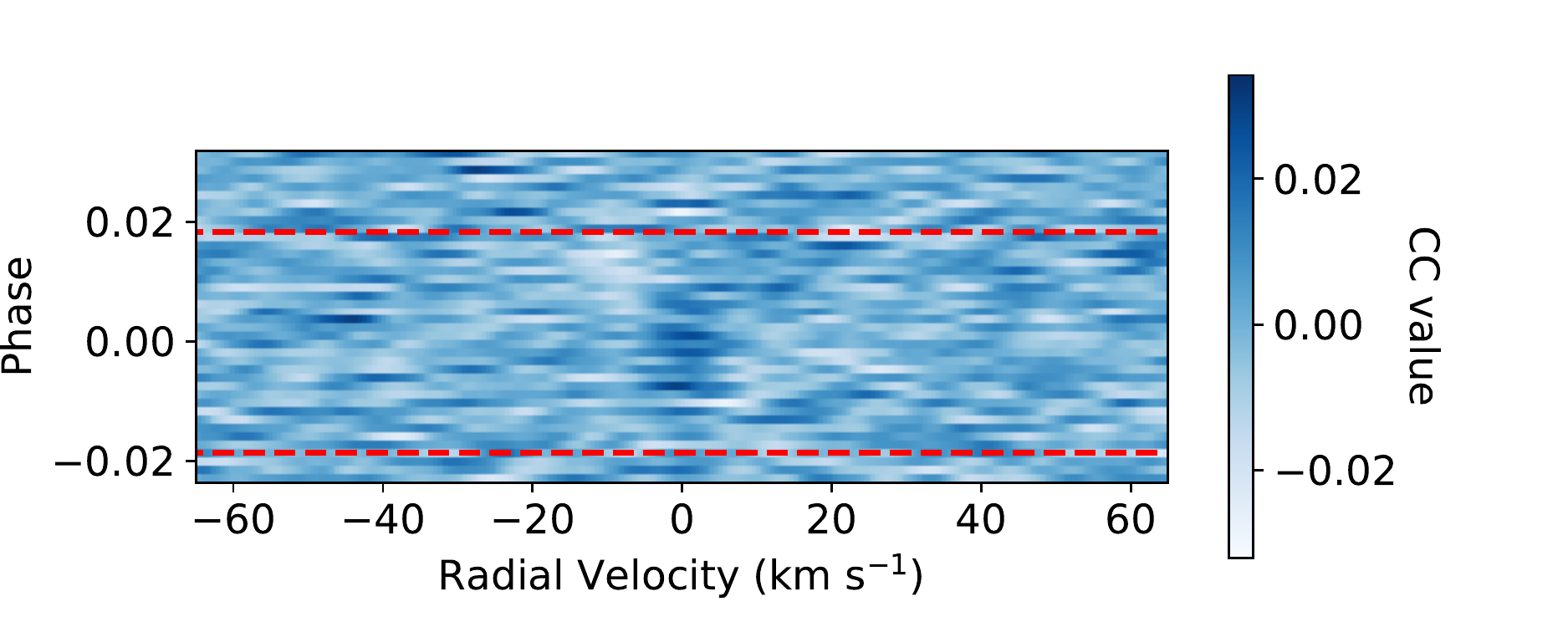}
\includegraphics[width=0.95\linewidth]{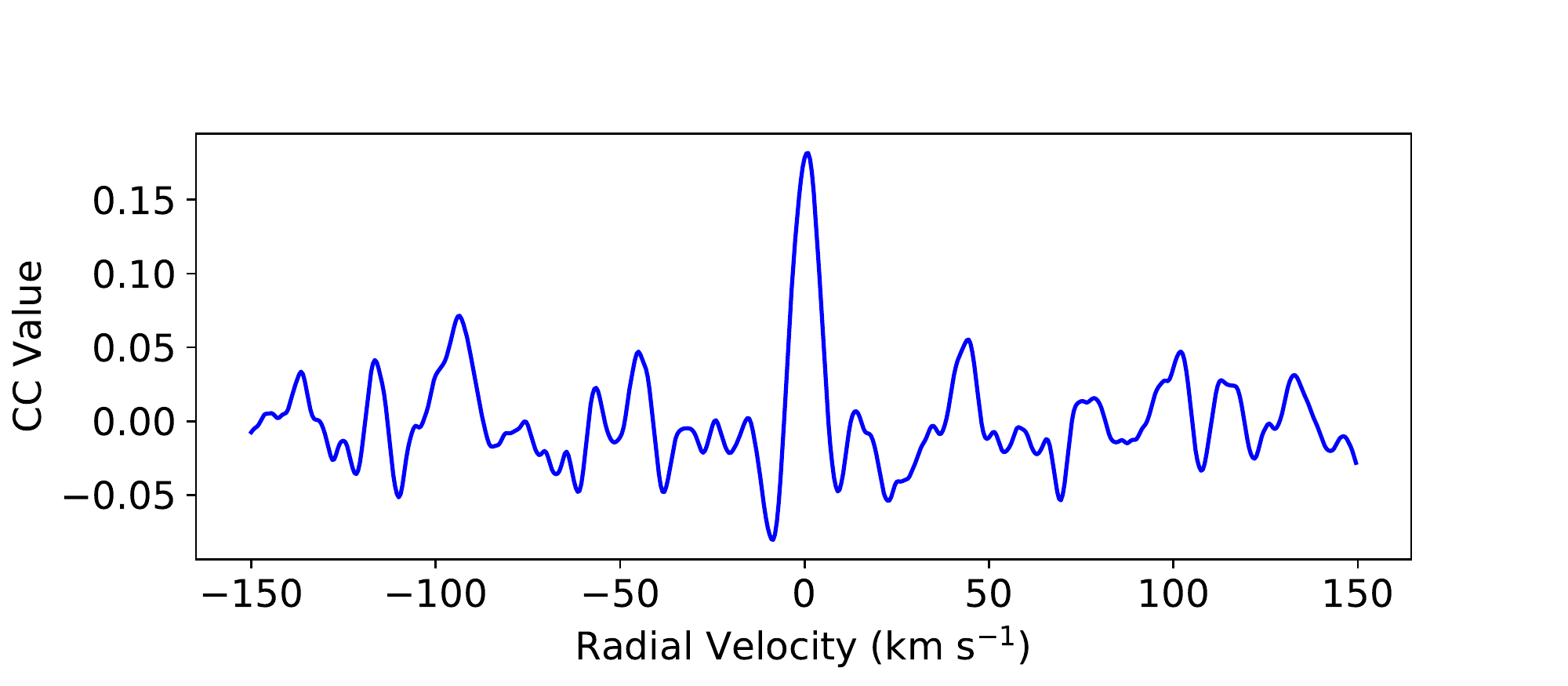}
\caption{\small 
Example cross correlation functions of HD 189733b with a model atmosphere. In all of the plots a planetary signal has been injected for demonstration purposes. The top panel shows all 46 cross correlation functions in the rest frame of the host star, without any \texttt{SYSREM} iterations. The red dashed lines indicate when the transit begins and ends, while the black dashed line indicates the planet's expected radial velocity. The signal appears strongest at zero phase due to the somewhat v-shaped transit of HD 189733b. The second panel (as well as the subsequent two) is the same as the top plot, except the optimal number of \texttt{SYSREM} iterations were applied during the analysis. The third panel shows the same as the second, except shifted into the planet's rest frame (the signal is now vertical and centered at 0 km s$^{-1}$). In the bottom panel all of the individual cross correlation functions have been weighted by the transit shape and then added together to create a single 1D cross correlation function. It is with this function that we can measure the SNR by computing the peak signal and the standard deviation of the surrounding noise. This injected signal has a SNR of 6.6, and so it would be considered a statistically significant detection.  }
\label{f:XcorlExample}
\end{center}
\end{figure}

To search for the exoplanetary signal in the data, each spectrum was cross correlated with the model for a wide range of radial velocities, spanning $-$250 to +250 km s$^{-1}$. Since we already interpolated the spectra onto a uniform grid in radial velocity space, no other steps were required to prepare the spectra for cross correlation. We normalized the cross correlation functions according to \citet{Tonry1979}. After the cross correlation, we were left with a grid containing a different cross correlation function for each time-series spectrum. 

The planet's velocity at the time of each spectrum can be calculated with the following equation 

\begin{equation} 
v_p(t, K_p) = v_{sys} + K_p \sin 2 \pi \phi(t)
\end{equation} 

\noindent where $v_{sys}$ is the systemic radial velocity of the star, $K_p$ is the semi-amplitude of the planet's radial velocity, and $\phi(t)$ is the orbital phase at the time of the observation. 
The values of $K_p$, $v_{sys}$, the time of transit $T_0$ and orbital period $P$ for each planet are given in Table \ref{t:planet1}.




Using the calculated $v_p$ values, we determined if there was a positive correlation between the model and data at the planet's expected velocity. The top panel of Figure \ref{f:XcorlExample} shows an example of the cross correlation matrix along with the calculated planet velocities. If the transmission spectrum shows significant FeH absorption, positive correlation should be present along the planet's velocity path. In Figure \ref{f:XcorlExample}, a signal was injected for clarity. To determine the strength of the signal and the resulting SNR of the possible detection, the cross correlation functions were each shifted to the planet's rest frame (third panel). We then added together all of the individual cross correlation functions, weighted by the transit depth at each phase (bottom panel). We implemented the PyTransit software package \citep{Parviainen2015} to model the transit light curve and determine the transit depth at the observed phases of each exoplanet, using a quadratic limb darkening model originally laid out in \citep{Mandel2002}. All the parameters used as input in the model are given in Table \ref{t:planet1} except for the limb darkening coefficients, which are from \citet{claret12, Claret2013}.  

\subsection{Injection and Recovery Tests}
\label{s:injections}

\begin{figure}
\begin{center}
\includegraphics[width=\linewidth]{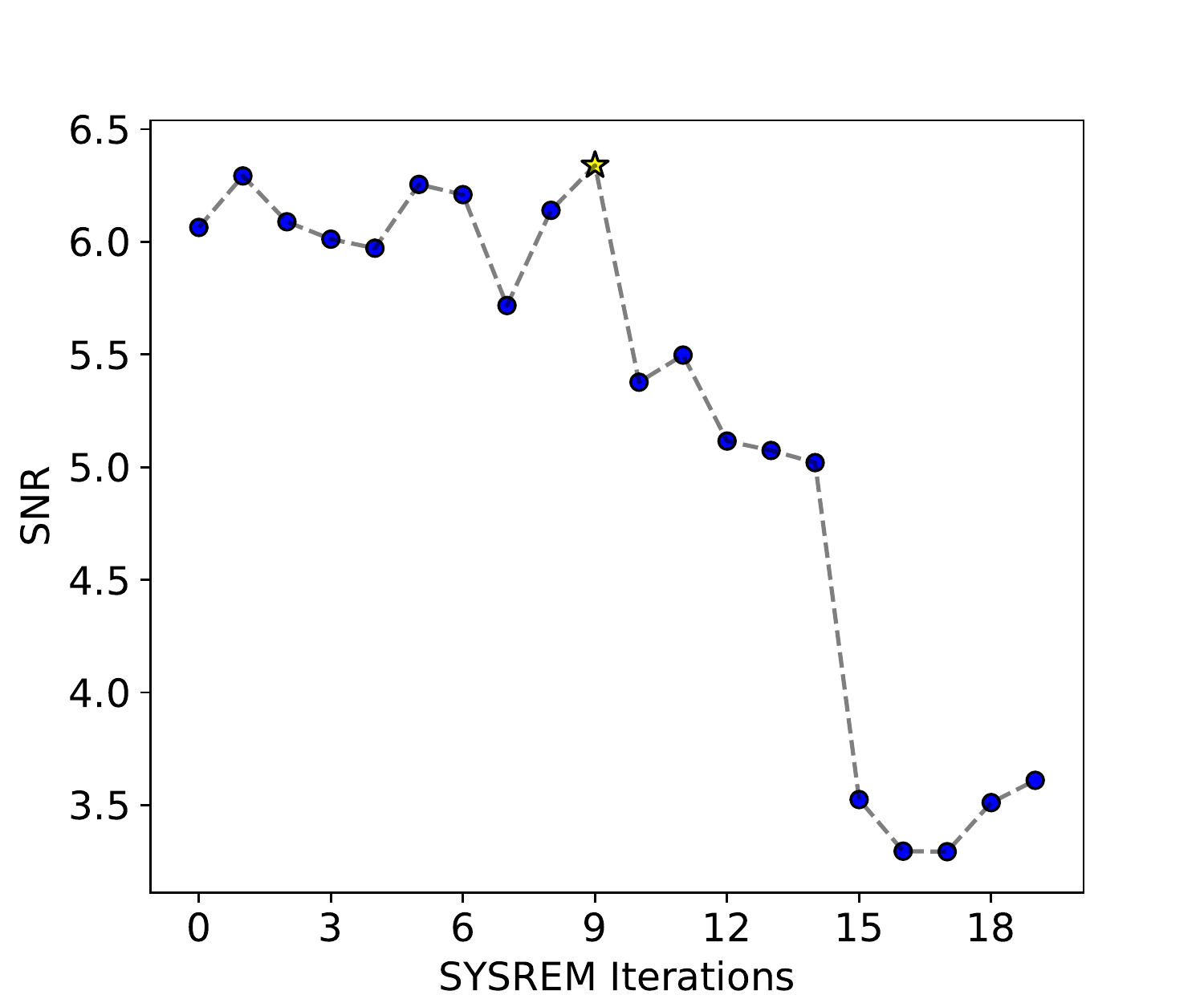}
\caption{\small 
Example plot demonstrating how the SNR changed as a function of \texttt{SYSREM} iteration. Again, we used HD 189733 as our example and have injected a signal with a VMR of 10$^{-7}$. Due to the region being relatively devoid of telluric lines, we recovered the signal with a SNR of 6.0 without any \texttt{SYSREM} iterations. However, by using \texttt{SYSREM} to remove the small telluric contamination, we were able to recover the signal at a higher SNR until nine iterations where \texttt{SYSREM} began to remove the planet's signal.  }
\label{f:sysrem}
\end{center}
\end{figure}

We performed a series of injection and recovery tests both to determine the optimal number of \texttt{SYSREM} iterations, and to determine the VMR of FeH that would be detectable in each planet's atmosphere. We will leave the discussion of the FeH VMRs for Section \ref{s:discussion}. 
To determine the optimal number of \texttt{SYSREM} iterations for each system, we injected a signal at the expected strength for a VMR that would be recoverable. We injected the signal at a different $v_{sys}$ value so as to not be biased by any real signal from the planet. The signal was injected at each phase with a strength specified by the PyTransit transit light curve, discussed in Section \ref{s:cc}. 
We then performed the full analysis as outlined in previous sections, changing only the number of \texttt{SYSREM} iterations. For each iteration we recorded the resulting SNR. Figure \ref{f:sysrem} shows an example of how the SNR changed with each \texttt{SYSREM} iteration. This analysis was performed separately for each night so that the optimal number of \texttt{SYSREM} iterations changed on a night-by-night basis. For the remaining analysis we used the number of \texttt{SYSREM} iterations that maximized the SNR. 

\section{Results} 
\label{s:results} 

\begin{figure*}
\begin{center}
\includegraphics[width=\linewidth]{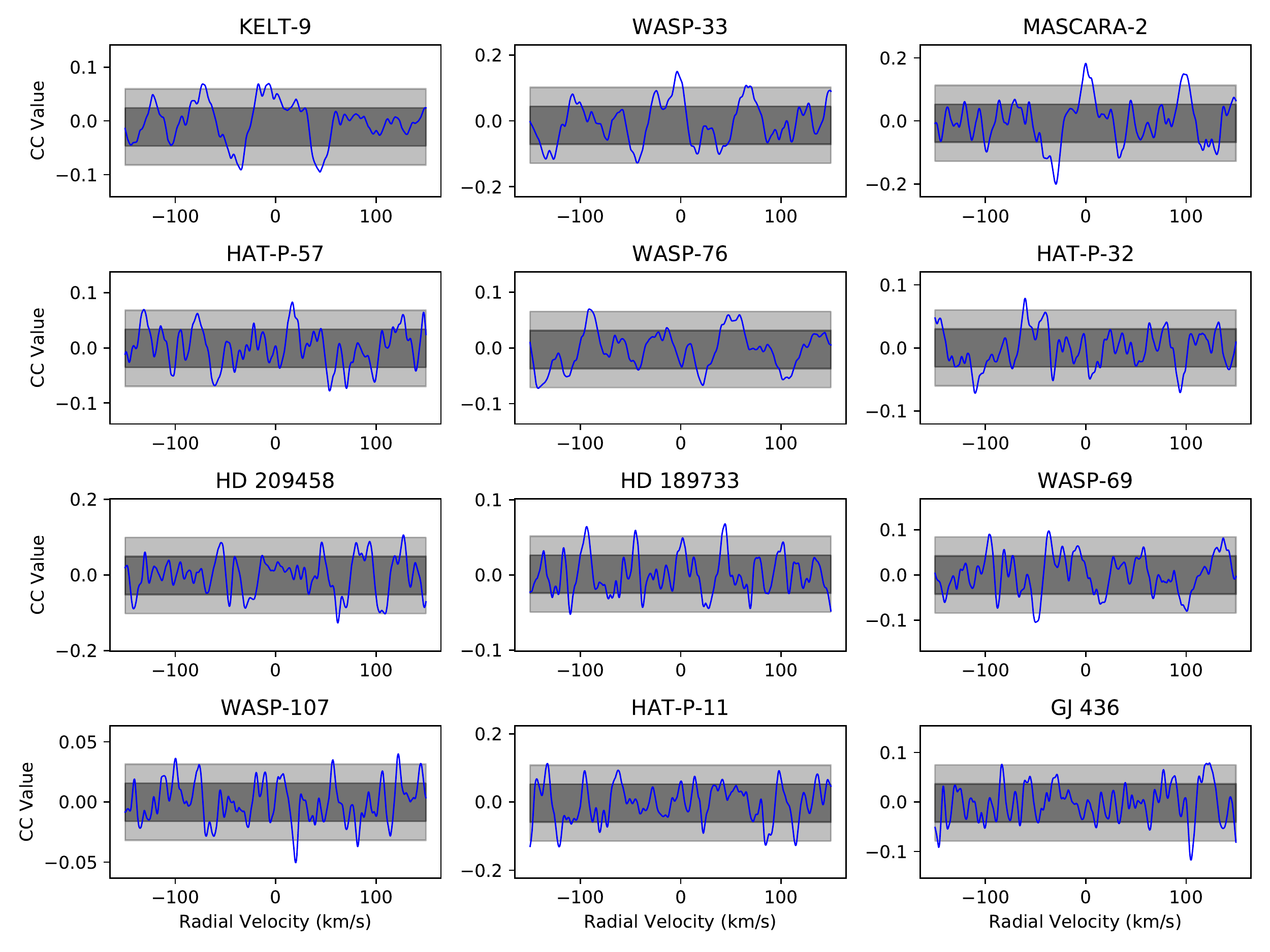}
\caption{\small 
One dimensional cross correlation functions between the FeH model atmosphere for each planet and the time series spectra. The spectra have all been shifted to the rest frame of the planet using the $K_p$ values in Table \ref{t:planet1}, and then added together over time in the same way as for Figure \ref{f:XcorlExample}. If FeH is present in a detectable amount in the planet's atmosphere a peak near 0 km s$^{-1}$ would be visible. The dark gray shaded region represents one standard deviation of the cross correlation function excluding a 30 km s$^{-1}$ region around zero, while the light gray region represents two standard deviations. WASP-33b and MASCARA-2b are the only planets with a peak above the two standard deviation near the expected location of 0 km s$^{-1}$. }
\label{f:flatXcorls}
\end{center}
\end{figure*}

We performed all of the analysis steps on each of the 12 exoplanet systems and searched for peaks in the cross correlation function along the planet's expected velocity. We did not find a significant detection of FeH in any of the exoplanets in our survey. Figure \ref{f:flatXcorls} shows all of the one dimensional cross correlation functions in the rest frame of the exoplanet. Two of the planets, WASP-33b and MASCARA-2b, show peaks near the planets' expected velocities at a SNR$\sim$3.
The peak of WASP-33b occurs at $-5$ km s$^{-1}$, while that of MASCARA-2b occurs at $-0.5$ km s$^{-1}$. Although these signals are not statistically significant, a slight offset in the system radial velocity is often seen in hot Jupiters, and is attributed to winds in the exoplanet's atmosphere \citep[$v_{wind}$;][]{Snellen2010, Ehrenreich2020}. 

By observing a larger sample of exoplanets, the probability of observing a 3-sigma peak due to random Gaussian noise in the expected velocity range increases. We performed a simple calculation to determine the probability of observing two of these 3-sigma peaks at the correct velocity. We assumed that any peak with wind velocities between 0 and -10 km s$^{-1}$ would be acceptable, which gives a window that is about 3 times the spectral resolution of CARMENES. The probability of measuring a random positive 3-sigma peak is 0.15\%. Therefore, with 12 exoplanets observing a 3-sigma noise peak within the expected velocity range has a probability of  $12\times3\times0.15\% = 5.4\%$, which combined for two of such objects gives 0.3\%. This is at about the 3$\sigma$ level, which, although interesting, we do not consider statistically significant enough. This 3-sigma level is a simple order of magnitude estimate and could be altered due to effects of correlated noise from tellurics or stellar residuals.

\begin{figure}
\begin{center}
\includegraphics[width= \linewidth]{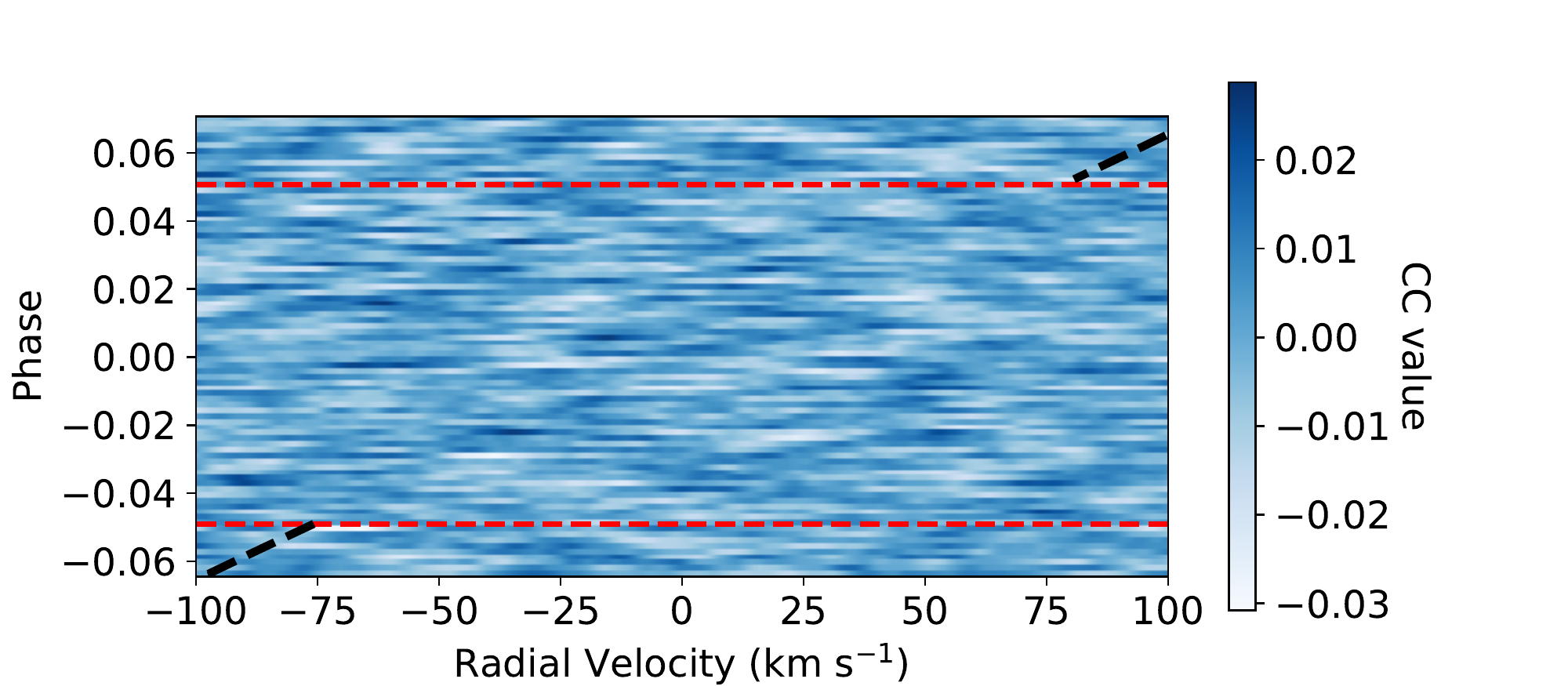}
\includegraphics[width= 0.95\linewidth]{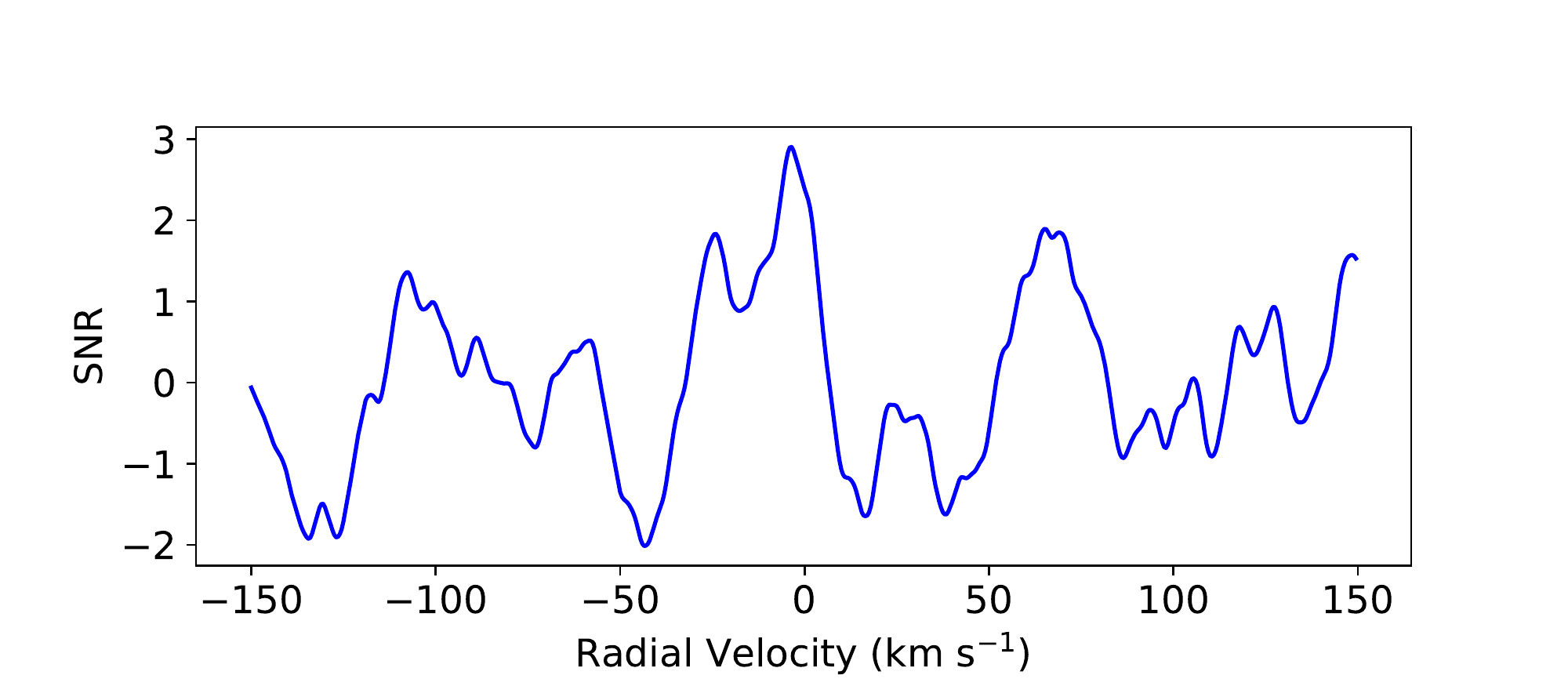}
\caption{\small 
Cross correlation matrix (top) and one dimensional cross correlation function (bottom) of WASP-33b after the 15 \texttt{SYSREM} iterations, using the optimal model and $K_p$ value of 248 km s$^{-1}$.  The one dimensional correlation function has a peak at -5 km s$^{-1}$ with a SNR of 2.98, which although interesting we do not consider to be statistically significant considering the size of our exoplanet sample.  }
\label{f:WASP33_Xcorl}
\end{center}
\end{figure}

\begin{figure}
\begin{center}

\includegraphics[width=\linewidth]{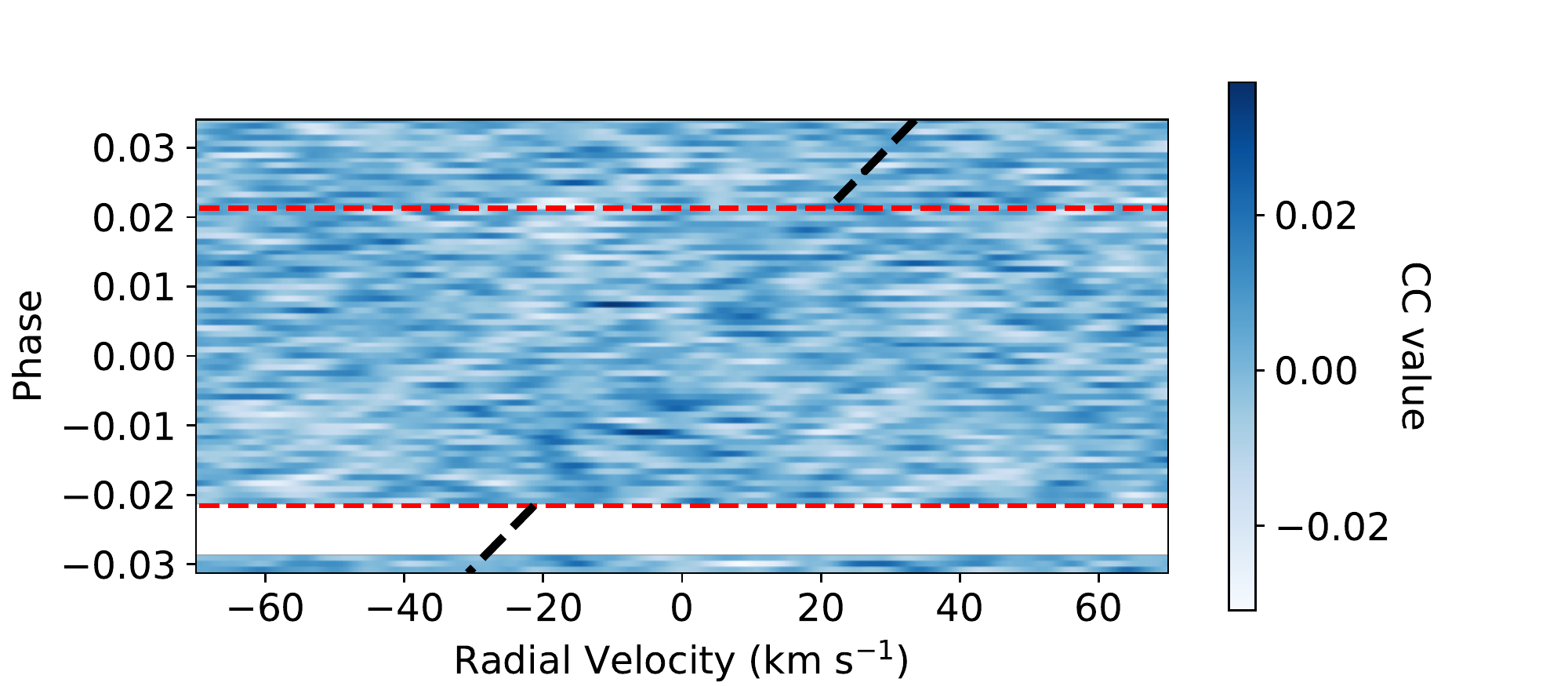}
\includegraphics[width= 0.95\linewidth]{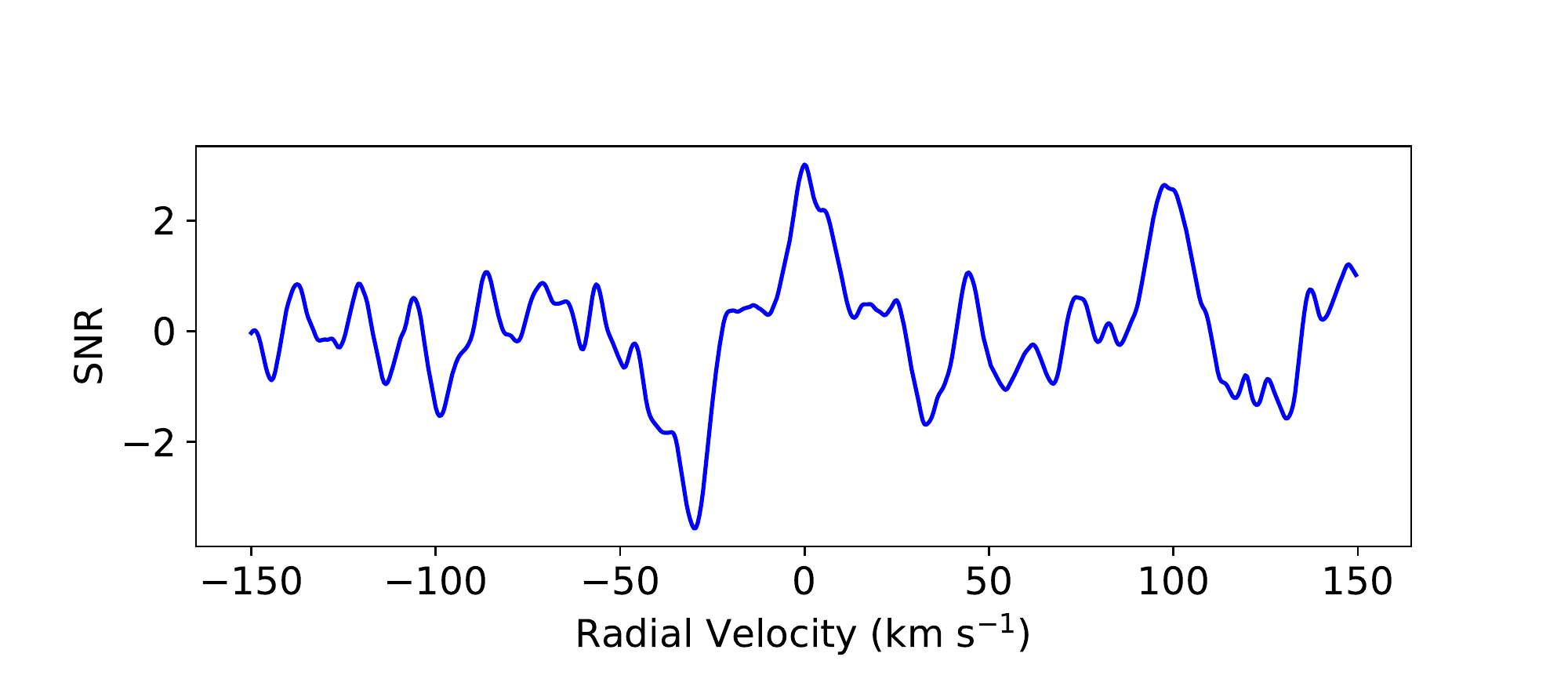}
\caption{\small 
Same as Figure \ref{f:WASP33_Xcorl} but for MASCARA-2b after 8 \texttt{SYSREM} iterations at a $K_p$ value of 158 km s$^{-1}$. The rows of white space are due to a gap in observing coverage, and no data exists on the archive directly before the start of the transit. The one dimensional correlation function has a peak at $-0.5$ km s$^{-1}$ with a SNR of 3.02, which although interesting we do not consider to be statistically significant considering the size of our exoplanet sample.  }
\label{f:MASC_Xcorl}
\end{center}
\end{figure}

We tested the validity of these signals further and explored whether they could be enhanced by small variations to the model or slightly different $K_p$ values. We found that for WASP-33b, the model that gave the largest SNR had a FeH VMR of 10$^{-9}$. For MASCARA-2b, a VMR of 10$^{-7}$ maximized the SNR. However, in both cases the change in SNR for volume mixing ratio differences of one order of magnitude (e.g., VMR of 10$^{-9}$ versus 10$^{-8}$) is roughly $\sim$0.2. In the following section we discuss whether values of 10$^{-7}$ to 10$^{-9}$ are reasonable for the VMR of FeH. Figures \ref{f:WASP33_Xcorl} and \ref{f:MASC_Xcorl} show the two dimensional and one dimensional cross correlation functions for both of these planets. 

\begin{figure*}
\begin{center}
\includegraphics[width= 0.48\linewidth]{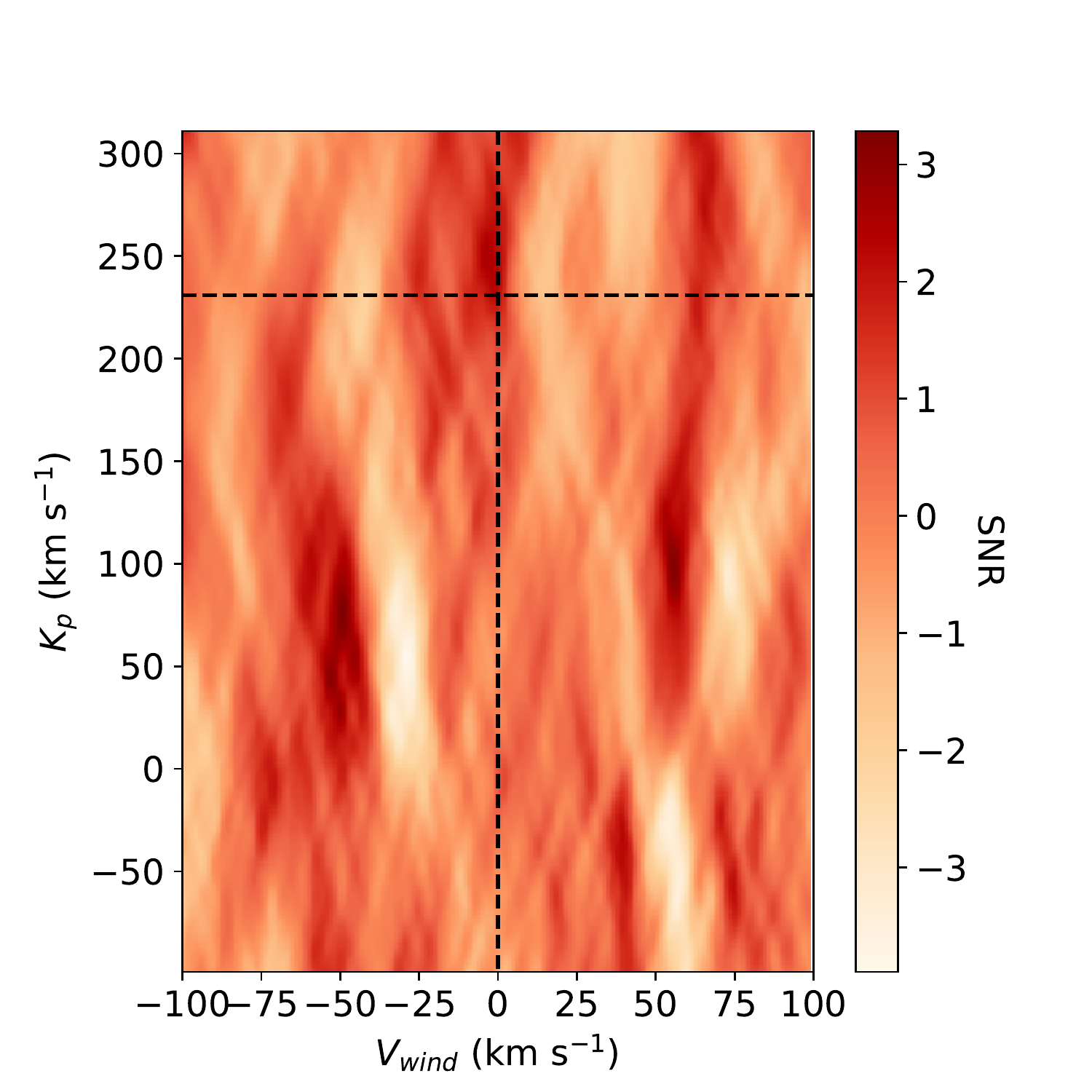}
\includegraphics[width= 0.48\linewidth]{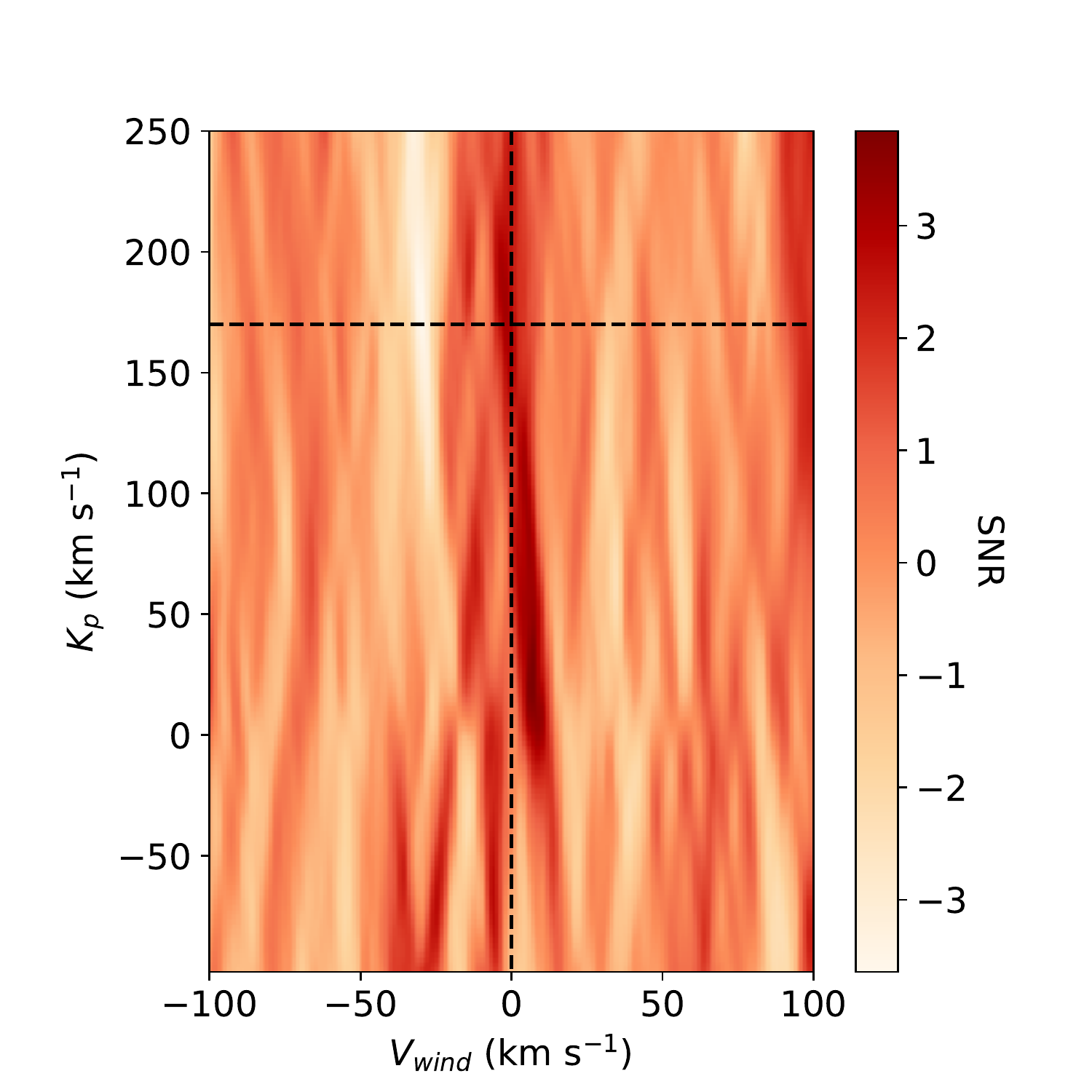}
\caption{\small 
Parameter space search for cross correlation peaks for a range of $K_p$ and $v_{wind}$ values in WASP-33b (left) and MASCARA-2b (right). The dotted black line shows the expected $K_p$ value for a $v_{wind}$ of 0 km s$^{-1}$. For WASP-33b, the peak in SNR near the expected value is not the strongest and is one of several peaks of similar SNR. The spurious strong peaks seem to be due to correlated noise in the cross correlation functions, which could be a signature of stellar pulsations.
For MASCARA-2b, other major SNR peaks reside near a $K_p$ and $v_{wind}$ of 0 km s$^{-1}$, leading us to believe they are associated with telluric contamination or stellar residuals. MASCARA-2b had the highest level of telluric contamination of all the spectra we analyzed, and we found that even with a large number of \texttt{SYSREM} iterations some telluric signal remained visible. The 0 km s$^{-1}$ peak seems to extend up toward the potential exoplanet signal peak, which further weakens the case for a true signal.}
\label{f:KpVsV}
\end{center}
\end{figure*}

We searched in a large region of $K_p$ and $v_{wind}$ parameter space to see if the assumed values represented the maximum signal, and to investigate any other strong features (see Figure \ref{f:KpVsV}). For WASP-33b, we find that the SNR $\sim$3 peak that occurs near the expected $K_p$ value has a maximum at $248\substack{+22\\-18}$ km s$^{-1}$. For MASCARA-2b we found that the maximum $K_p$ occurred at $158\substack{+41\\-14}$ km s$^{-1}$. These uncertainties represent the $K_p$ values where the SNR decreases by one for the peak near the planet's expected velocity. We note that these are not one sigma error bars as a decrease in one of SNR does not necessarily directly correspond to a decrease of one in sigma.  

Neither parameter search (Figure \ref{f:KpVsV}) convincingly shows a detection of FeH as both plots reveal other strong positive correlation peaks that can certainly not be associated with the planet. WASP-33b is a known Delta Scuti pulsator, and while we do not see obvious residuals from the pulsations in our analysis, some of the noisy areas in the 2D cross correlation functions could be due to residuals from the removal of stellar lines that were affected by the pulsations (see area between 25 and 75 km s$^{-1}$ in Figure \ref{f:WASP33_Xcorl}). A previous analysis of the same WASP-33 data by \citet{Yan2019} found that signals from Ca could be successfully recovered from the data after applying a high-pass filter similar to the one applied in our analysis, but they noted that a particularly strong negative correlation signal that still remained in their data could be due to the pulsations. Alternatively, the spurious signals in Figure \ref{f:KpVsV} for WASP-33 could simply be due to the low SNR of the data. Other peaks in the plot of MASCARA-2b seem to be caused by telluric contamination, stellar residuals, or some combination of the two, as the SNR peak extends down to 0 km s$^{-1}$. Therefore, even though the 1D cross correlation functions show SNRs $\sim3$, we do not claim a statistically significant FeH detection. 

Some previous studies have questioned the statistical significance of solely using the SNR metric to judge the quality of the signal \citep[e.g.,][]{Brogi2013}. We therefore also tested the use of the Welch T-test to compare the signal within and outside the expected exoplanet's trail, as has been done in many previous works \citep[e.g.,][]{Brogi2013, Alonso2019a}. We find that the Welch T-test gives a very similar statistical significance for the potential signal in WASP-33b, but that it estimates an extremely high significance of approximately 5$\sigma$ for the signal in MASCARA-2b. The Welch T-test has been previously shown to often lead to over-inflated confidence estimates \citep{Cabot2019}. Furthermore, the test does not take into account any correlated noise, and so we are skeptical of these results and instead prefer to report the significance of the MASCARA-2b detection with a SNR of 3.

\section{Discussion}
\label{s:discussion} 

\begin{figure*}
\begin{center}
\includegraphics[width=\linewidth]{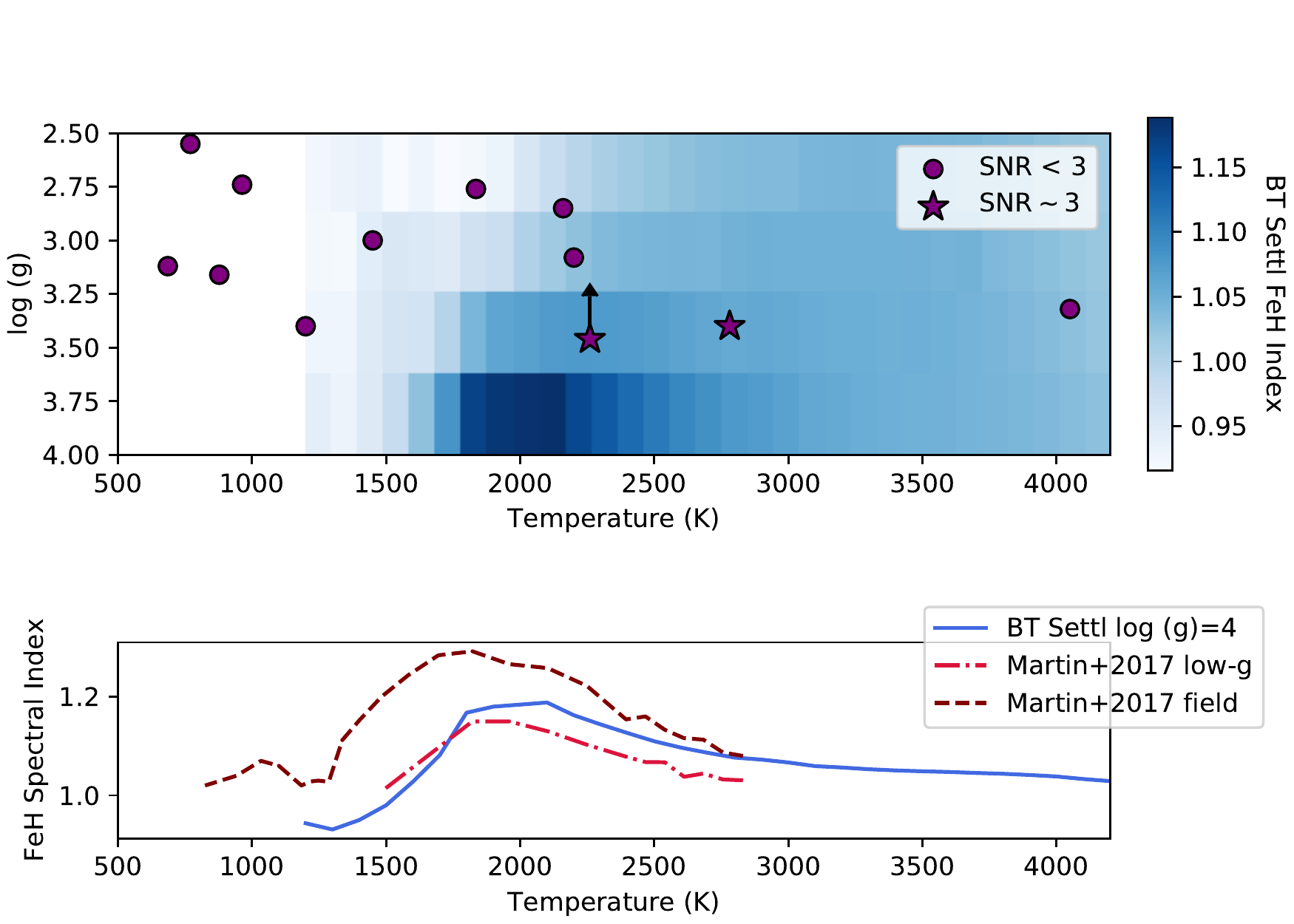}
\caption{\small  
\textbf{Top:} Expected FeH feature strength over the parameter space explored in this paper. Each symbol represents a planet in our study, where the temperatures are the equilibrium temperatures of the planet. The purple circles are planets where we do not detect any FeH, while the stars indicate the two $\sim3-\sigma$ signals (WASP-33b and MASCARA-2b). The blue colormap shows the expected FeH feature strength (darker blue is stronger FeH absorption), calculated by measuring its spectral index from a grid of BT-Settl models with a given effective temperature and surface gravity \citep{Allard2012}. Although we do not consider the two $\sim3-\sigma$ signals statistically significant, it is interesting that they lie in the part of parameter space with strong expected FeH absorption. \textbf{Bottom:} Comparison between the FeH spectral index calculated from the models with a log $g$ of 4.0 and FeH spectral indices of brown dwarfs with similar temperatures from \citet{Martin2017}.  The red dot-dashed line is the trend for low gravity brown dwarfs (3.5$\gtrsim$log $g$ $\gtrsim$4.5), while the maroon dashed line is for field brown dwarfs (log $g\gtrsim$5.0).  The models seem to be in good agreement with the low-gravity data, and the trend of stronger FeH lines with higher log $g$ values is present in both the models (blue scale in top panel) and observations. }
\label{f:FeH_grid}
\end{center}
\end{figure*}

\begin{figure}
\begin{center}
\includegraphics[width=\linewidth]{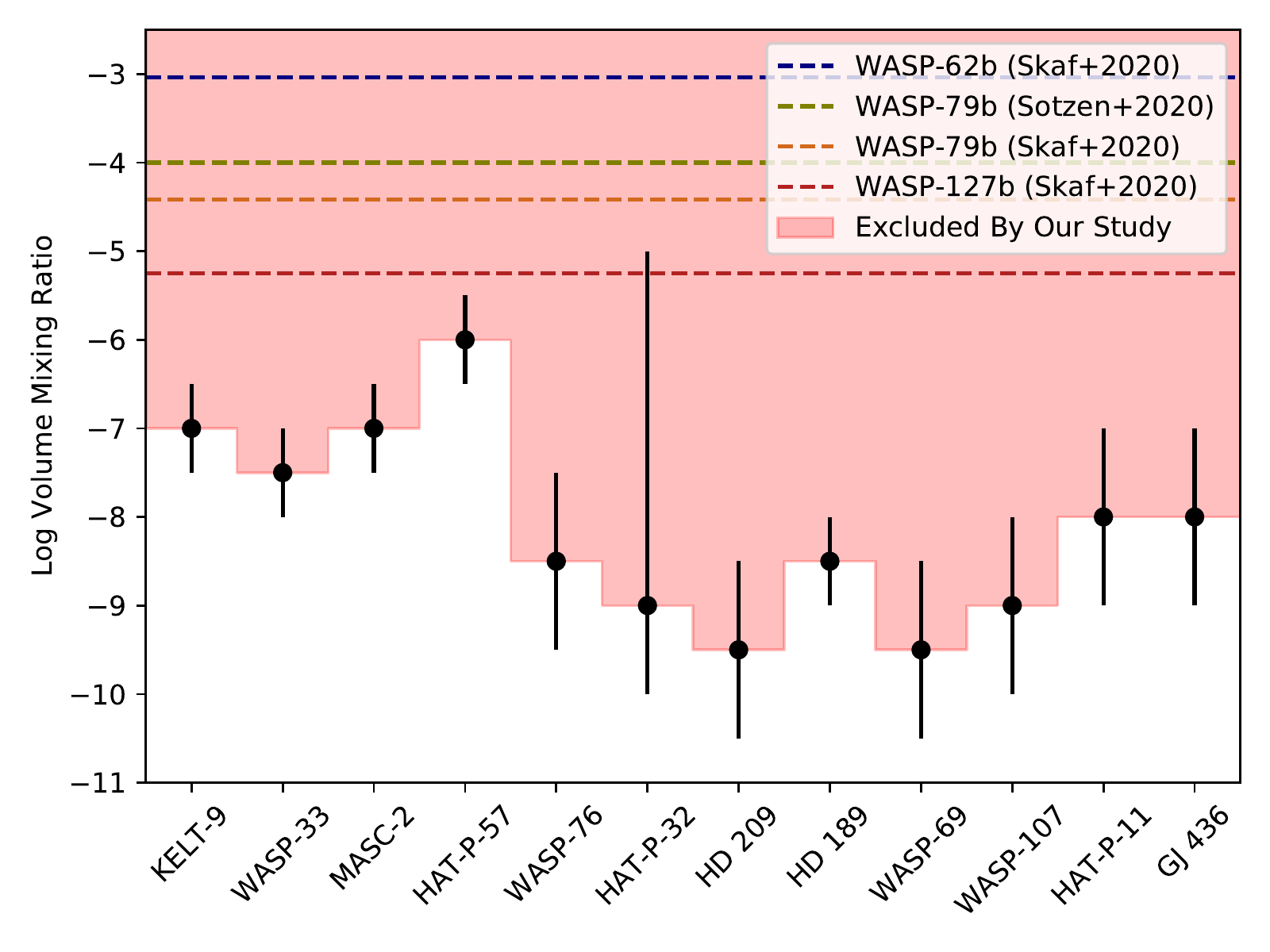}
\caption{\small  
Results of our injection and recovery tests for each planet. The black points represent the FeH VMR that we are able to recover with an SNR$>5$ for the fiducial model atmosphere parameters (bolded values in Table \ref{t:model}). The errors originate from the uncertainties in the atmospheric model parameters, such as the location of the cloud base. The light red shaded region shows VMRs that we would successfully be able to recover with the data presented in this paper. The dotted lines shows the VMR of FeH that \citet{Sotzen2020} and \citet{Skaf2020} recover for WASP-62b, WASP-79b, and WASP-127b. If FeH was present in any of the recovered abundances we would be able to detect it in every planet in our sample.   }
\label{f:injection}
\end{center}
\end{figure}

We searched for signals of FeH in 12 exoplanets spanning a large range of equilibrium temperatures and log $g$, but found no conclusive detections of FeH. In two of the exoplanets, however, we did see SNR $\sim$ 3 signals at the expected planet's velocity. In order to interpret the results, it is important to discuss what the VMR of FeH is expected to be over the range of parameter space studied here, and what VMR of FeH we would be able to recover for each planet. To explore where in temperature and pressure parameter space FeH is expected to be most abundant, we calculated the strength of the FeH main bandhead in a grid of PHOENIX BT-Settl atmosphere models \citep{Allard2012, Baraffe2015}. We measured the FeH spectral index for each model spectrum by simply calculating the average continuum flux directly before the start of the FeH bandhead ($9850 - 9895$\AA) divided by the average flux within the FeH bandhead ($9900 - 10000$\AA). The top panel of Figure \ref{f:FeH_grid} shows the planets in our study plotted over the FeH spectral indices from the models. 

By using an FeH spectral index to gauge the FeH strength, we were also able to compare the models to observations of brown dwarfs with similar temperatures, since FeH spectral indices are commonly reported for these objects. The bottom panel of Figure \ref{f:FeH_grid} shows this comparison between models with a log $g$ of 4.0 and observations from \citet{Martin2017}. The spectral indices in \citet{Martin2017} were originally reported in spectral types, which we converted into temperatures with the relation from \cite{Filippazzo2015}. Metal hydrides, such as FeH, have long been known to be gravity dependent, and \citet{Martin2017} used the J-band FeH spectral index along with other atomic lines to divide a large sample of brown dwarfs into surface gravity bins. The FeH spectral indices from the low-gravity brown dwarfs match those from the models very well. \citet{Martin2017} also measured that the FeH spectral indices from the higher gravity objects were larger, helping to validate the trend seen in the top panel of Figure \ref{f:FeH_grid}.

The other trend that is visible in both panels of Figure \ref{f:FeH_grid} is that FeH exhibits the strongest absorption between about 1800 and 3000 K. 
Below temperatures of about 1800 K iron starts to condense out of the atmosphere and the abundance quickly drops off \citep{Visscher2010}. At higher temperatures, FeH dissociates and again the abundance decreases. The observations of brown dwarfs show that FeH again becomes visible at temperatures of around 1000 K. It is not exactly known why FeH becomes visible again, but \citet{Burgasser2002} suggested that it could be evidence of cloud disruption, which allows deeper layers of the atmosphere to be probed. This suggests that in future studies, FeH could be a tool for uncovering weather and cloud dispersal in planetary atmospheres at low temperatures.

The two objects in our sample for which we detect a CCF peak near the expected $K_p$ and $V_{sys}$ have equilibrium temperatures and log $g$ values corresponding to where FeH is expected to produce a strong signal. WASP-33b and MASCARA-2b have high enough log $g$ values that FeH is still quite abundant, but not too high that the scale height is so small that the atmosphere cannot be effectively probed with transmission spectroscopy. Since the highest FeH abundance is expected for high gravity objects, day-side spectroscopy may more amenable to FeH searches due to fact that the signal strength does not decrease for smaller scale heights as it does in transmission spectroscopy. Furthermore, emission spectroscopy can probe deeper in the atmosphere, where FeH is thought to be more abundant \citep{Visscher2010}.

\citet{Visscher2010} studied the chemical behavior of iron-bearing gases in giant planets, brown dwarfs and low-mass stars to derive abundances as a function of temperature, pressure, and metallicity. They found that FeH is the second most abundant iron-bearing gas after monatomic Fe at temperatures above about 1500 K. For these temperatures, and pressures between 0 and 10$^{-5}$ bars (the region of the atmosphere probed by transmission spectroscopy of most  molecules), the FeH abundance was found to be between 10$^{-7}$ and 10$^{-11}$. 

We used injection and recovery tests to determine what VMR of FeH we could recover for each planet. We used a SNR of 5 as our detection threshold. Figure \ref{f:injection} shows the FeH VMR limits for each planet for the range of different model atmospheres shown in Table \ref{t:model}. For all the planets we could recover abundances down to 10$^{-6}$ and for some as low as 10$^{-9.5}$. When we inject an FeH signal with a VMR of 10$^{-8}$ into the spectra of MASCARA-2b and a VMR of 10$^{-9}$ into WASP-33b, we recover both signals with a SNR of about 3. Although we do not treat the observed signals as statistically significant, their amplitudes are in the range for the expected VMRs. In addition, it indicates that obtaining a statistically significant detection of FeH in hot Jupiters requires transmission spectra at a signal to noise of about a factor of two better than what are currently available, but well within the capabilities of current facilities and instruments. 

Two studies have recently published potential FeH detections in three different exoplanets, WASP-79b, WASP-127b, and WASP-62b \citep{Sotzen2020, Skaf2020}. These are all hot Jupiters with equilibrium temperatures between 1300 and 1700 K and log $g$ values less than 2.9, which means that they have both lower surface gravities and lower temperatures than the planets for which we obtained potential signals. \citet{Sotzen2020} found a best fit FeH VMR of $\sim10^{-4}$ in WASP-79b, while \citet{Skaf2020} retrieved VMRs of $10^{-4.42}$, $10^{-5.25}$, and $10^{-3.04}$ from WASP-79b, WASP-127b and WASP-62b, respectively. These retrieved FeH VMRs are between three and five orders of magnitude more abundant than \citet{Visscher2010} predicted. If FeH exists at this level in any of the planets in our sample we would have detected it (see Figure \ref{f:injection}). While none of these planets are in our sample, the orders of magnitude discrepancy between the potential FeH VMRs is worrying and could point to over-estimation of molecular opacities in low-resolution data due to degeneracies with clouds, hazes, or H$^{-}$ continuum opacity. 

Using both low- and high-resolution observations leads to a more complete and accurate picture of the role of metal hydrides and oxides in hot-Jupiter atmospheres. This combination of low- and high-resolution studies has proved vital for uncovering whether TiO and VO are present in WASP-121b \citep[][]{Evans2016, Evans2018, Merritt2020}, and highlights the importance of obtaining both types of data.

While all of these methods for estimating the VMR of FeH in exoplanetary atmospheres can give us a general idea of how the chemical abundances change with temperature and surface gravity, it is important to note that they may not give a full picture, and important physics still may be missing. Atmospheric models and low-gravity brown dwarfs both address the issue that planets have lower log $g$ values than stars and field brown dwarfs, but neither have the intense insolation from the host star, which can cause ultra-hot Jupiters to host temperature inversions \citep{Pino2020} and potentially change the predicted VMRs of FeH. This intense insolation also causes large day-to-night temperature constrasts, and iron could be rained out on the night side of hot Jupiters, as seen in WASP-76b \citep{Ehrenreich2020}, which would have unknown affects on the abundance of FeH on the day side and terminators. Additionally, the chemical models describing the behavior of iron in \citet{Visscher2010} assume thermochemical equilibrium, which is not always an accurate assumption, especially in the upper parts of the atmosphere that are probed by transmission spectroscopy \citep{Molaverdikhani2019}. Because of these issues, measuring FeH in a variety of exoplanetary atmospheres will be needed to fully understand how these difference affect its abundance. 

\section{Conclusions\label{s:conclusions} } 

We searched for FeH in archival near-infrared 
CAR-MENES spectra of 12 exoplanets spanning a wide range in temperature and surface gravity. The FeH main bandhead is located at 0.99 $\mu$m, which is ideally situated between two water bands, making it relatively free from telluric contamination. Since removing tellurics is the most challenging part of ground-based high-dispersion transmission spectroscopy, the location of the bandhead is ideal to efficiently and accurately search for FeH in a large sample of planets. To search for the exoplanet's FeH signal we cross correlated the data with a range of exoplanet atmospheric models, created using petitRADTRANS. 

We did not find any statistically significant FeH signals in any of the transmission spectra. Two of the planets, WASP-33b and MASCARA-2b, showed positive correlation near the expected $K_p$ and $v_{sys}$ with SNRs of about 3. Even though these peaks seemed promising in the 1D cross correlation functions, there were several other peaks with similar or greater significance when we searched a wider $K_p$ and $v_{sys}$ range, and so we do not claim a detection. 

To put these results into context, we explored what the expected VMR of FeH would be for each planet, and where in parameter space FeH would contribute the most opacity. We conclude that opacity from FeH is most likely important for planets with temperatures between 1700 and 3000 K, and relatively high log $g$ values. However, at lower temperatures FeH could still be important if clouds are somehow dispersed. While the signals of WASP-33b and MASCARA-2b are not statistically significant, it is interesting that these two planets reside in the part of parameter space where the expected FeH opacity is strong. 

By performing injection and recovery tests we were able to rule out FeH existing in any of these exoplanets' atmospheres with a VMR greater than $10^{-6}$ and for some, as low as VMRs of $10^{-9.5}$. If WASP-33b contained FeH with a VMR of $10^{-9}$ and MASCARA-2b a VMR of $10^{-8}$, our injection and recovery tests indicate that we would recover a SNR of about 3, similar to what we extract from the data. Chemical modeling of iron in planetary atmospheres suggests that the FeH VMR is most likely between $10^{-7}$ and $10^{-10}$. Recent results from HST transmission spectra retrieve much higher FeH VMRs (between $10^{-3}$ and $10^{-5}$), which is at odds with our results and those from chemical modeling, highlighting the importance of high-resolution data. 

We conclude that FeH could potentially exist in the atmospheres of WASP-33b and MASCARA-2b at a VMR of $10^{-8} - 10^{-9}$, but that higher quality data or more transits is required to reject or confirm the planetary nature of these signals. Measurements of FeH in hot-Jupiter atmospheres is therefore well within the observing limits of future ground-based high-dispersion spectroscopy studies. 

\acknowledgements

A.K., I.S., and D.S. acknowledge funding from the European Research Council (ERC) under the European Union's Horizon 2020 research and innovation program under grant agreement No 694513. P.M. acknowledges support from the European Research Council under the European Union's Horizon 2020 research and innovation program under grant agreement No. 832428. We would like to thank Alex Cridland and Alejandro Sanchez-Lopez for useful discussions during the preparation of the manuscript. We would also like to thank the CARMENES consortium for making their data publicly available. This research has made use of the NASA Exoplanet Archive, which is operated by the California Institute of Technology, under contract with the National Aeronautics and Space Administration under the Exoplanet Exploration Program. 

\facilities{CAO:3.5m (CARMENES)}

\software{\texttt{astropy} \citep{astropy}, \texttt{matplotlib} \citep{matplotlib}, \texttt{numpy} \citep{numpy}}

\bibliographystyle{aasjournal}
\bibliography{bib.bib}

\end{document}